\documentclass[english,aps,prl,twocolumn,graphics,graphicx]{revtex4-1}
\usepackage[utf8]{inputenc}

\usepackage{amsmath,mathtools}
\usepackage{amssymb}
\usepackage{graphicx}
\usepackage{verbatim}
\usepackage{braket}
\usepackage{soul}
\usepackage{bm}
\usepackage{gensymb}
\usepackage{diagbox}
\usepackage{siunitx}
\usepackage[T1]{fontenc}
\usepackage[english]{babel}
\usepackage{mathtools}
\usepackage{newtxtext}  
\usepackage{amsmath}
\usepackage{amsfonts}
\usepackage{hyperref}

\renewcommand\vec{\boldsymbol}

\begin{document}
\title{Influence of the Dirac Sea on Phase Transitions in Monolayer Graphene under Strong Magnetic Fields}


\author{Guopeng Xu}
\author{Chunli Huang}
\affiliation{Department of Physics and Astronomy, University of Kentucky, Lexington, Kentucky 40506-0055, USA}
\date{\today} 

\begin{abstract}

Recent scanning tunneling microscopy experiments have found Kekul\'e-Distorted (KD) ordering in graphene subjected to strong magnetic fields, a departure from the antiferromagnetic (AF) state identified in earlier transport experiments on double-encapsulated devices with larger dielectric screening constant $\epsilon$. This variation suggests that the magnetic anisotropic energy is sensitive to dielectric screening constant.
To calculate the magnetic anisotropic energy without resorting to perturbation theory, we adopted a two-step approach. 
First, we derived the bare valley-sublattice dependent interaction coupling constants from microscopic calculations and account for the leading logarithmic divergences arising from quantum fluctuations by solving renormalization group flow equations in the absence of magnetic field from the carbon lattice scale up to the much larger magnetic length. Subsequently, we used these renormalized coupling constants to perform non-perturbative, self-consistent Hartree-Fock calculations. Our results demonstrate that the ground state at neutrality ($\nu=0$) transitions from a AF state to a spin-singlet KD state when dielectric screening and magnetic fields become small, consistent with experimental observations. For filling fraction $\nu=\pm1$, we predict a  transitions from spin-polarized charge-density wave states to spin-polarized KD state when dielectric screening and magnetic fields become small. Our self-consistent Hartree-Fock calculations, which encompass a large number of Landau levels, reveal that the magnetic anisotropic energy receives substantial contributions from the Dirac sea when $\epsilon$ is small. Our work provides insights into how the Dirac sea, which contributes to one electron per graphene unit cell, affects the small magnetic anisotropic energy in graphene.

\end{abstract}
\maketitle

\section{Introduction}

\begin{figure*}[t]
    \centering
   \includegraphics[width=1.0\linewidth]{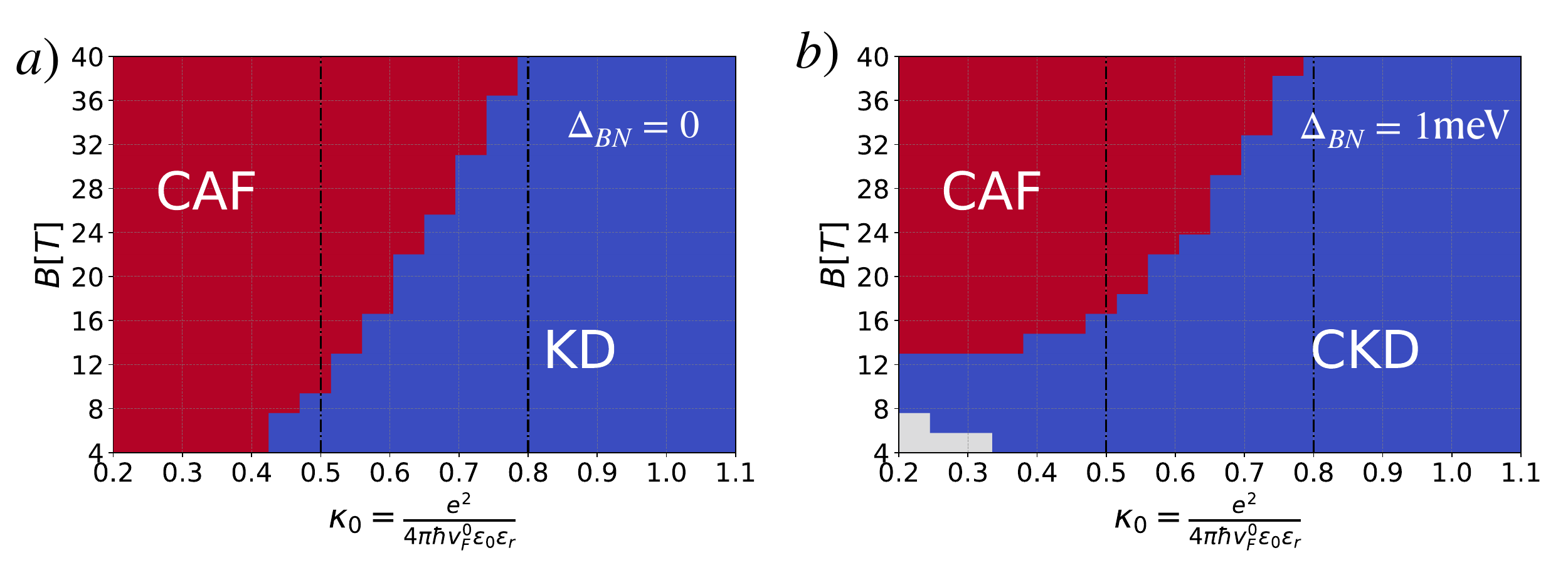}
    \caption{Zero temperature phase diagram of $\nu=0$ graphene computed with renormalized Fermi velocity and renormalized short-range interactions, incorporating both zeroth-Landau level and Dirac sea energy contributions.  It shows a first-order phase transition between the canted-antiferromagnet (CAF) and spin-singlet Kekul\'e-Distorted (KD) states. In panel a), the sublattice potential is zero while in panel b) a finite sublattice potential $\Delta_{BN}=1$
     meV induces sublattice polarization, resulting in the canted Kekul\'e-Distorted (cKD) state. The two vertical dashed lines denote the fine-structure constants for a double encapsulated sample ($\kappa_0=0.5$, $\epsilon_r=4$) and an open-surface sample ($\kappa_0=0.8$, $\epsilon_r=2.7$). Landau-level mixing effects intensify in the bottom right corner, where $\kappa_0$ is large and $B$ is small, favoring the KD state. The grey area indicates a BN -field-induced sublattice polarized state.}
    \label{fig:zero_LL_phase_diagram}
\end{figure*}

The exploration of quantum Hall ground states in monolayer graphene, enriched by its spin and valley degrees of freedom, has been the focus of intensive study over the past two decades\cite{PhysRevB.74.075422, PhysRevLett.96.256602,PhysRevB.74.075423,PhysRevB.79.115434,PhysRevLett.103.216801,Dean_2011,RevModPhys.83.1193,PhysRevLett.112.126804,doi:10.1126/science.aan8458,Zibrov_2018,Zhou_2019,Veyrat_2020,farahi2023brokensymmetriesexcitationspectra,doi:10.1126/science.adf9728,PhysRevLett.111.266801,PhysRevB.94.245435}  and continues to produce surprising discoveries \cite{Coissard2021ImagingTQ,doi:10.1126/science.abm3770,PhysRevB.100.085437, delagrange2024vanishingbulkheatflow,kumar2024absenceheatflownu,Zhou2021StrongMagneticFieldMT,doi:10.1126/science.aar4061,Stepanov2018LongdistanceST,PhysRevX.11.021012,PhysRevLett.126.117203}. Recent atomic-scale tunneling experiments\cite{Coissard2021ImagingTQ,doi:10.1126/science.abm3770,PhysRevB.100.085437} have demonstrated that neutral graphene can exhibit various phases, including spin-ordered phases, sublattice-polarized phases (sometimes referred to as charge-density waves), and Kekul\'e distorted (KD) phases (which triple the unit cell area of graphene), depending on the screening environment. Earlier non-local magnon transport experiments\cite{Zhou2021StrongMagneticFieldMT,doi:10.1126/science.aar4061,Stepanov2018LongdistanceST,PhysRevX.11.021012,PhysRevLett.126.117203} primarily identified the substrate-induced sublattice polarized  and canted-antiferromagnetic phases, but did not report the finding of Kekul\'e distorted (KD)  phases. 
Recent heat transport experiments report a vanishing bulk thermal conductance at neutrality, suggesting the absence of low-lying gapless (collective) modes in the symmetry-broken phase \cite{delagrange2024vanishingbulkheatflow,kumar2024absenceheatflownu}. 

When non-relativistic electron gases are subjected to strong magnetic fields, the first step in understanding the organization of strongly-correlated electrons typically involves projecting the many-body Hamiltonian onto specific Landau levels. However, this approach becomes less straightforward for relativistic Landau levels.  Directly projecting the many-body Hamiltonian onto a desired Landau level often proves unreliable, even under strong magnetic fields, because of the following reasons.
Firstly, the matrix elements governing Landau-level mixing are proportional to the fine structure constant of graphene, $\kappa_0 = e^2/(\hbar v_F^0 \epsilon_0\epsilon_r)$.
In devices where graphene is encapsulated by Boron-Nitride on both sides, with a screening constant $\epsilon_r=4.4$ and a bare Fermi velocity $v_F^0=1 \times 10^6$m/s -- the resulting $\kappa_0=0.5$ is not a small number. Secondly, the density of states of $\pi$ band increases with the energy relative to the Dirac point.  These two factors, arising from graphene's linear dispersion, together imply that remote Landau levels can substantially influence the Landau-level splittings close to the Dirac point, thus impacting the ground-state-order parameter.
Indeed, our second-order perturbation theory calculations in $\nu=0$ suggest that renormalized interactions tend to favor the Kekul\'e distorted state over the canted-antiferromagnetic state \cite{wei2024landau}. This tilt in balance arises from specific double-exchange Feynman diagrams and is mainly influenced by states deep within the Dirac sea, independent of the band structure details near the Dirac point. However, that analysis assumes $\kappa_0$ to be small and neglects the spin-valley polarization of the Dirac sea, whose contributions to the magnetic anisotropy are of higher order in $\kappa_0$ but can become important for realistic values of $\kappa_0$\cite{PhysRevB.80.235417}.

This uncertainty motivates us to develop a more comprehensive microscopic theory that moves beyond perturbation theory, addressing the complex interaction effects in graphene through a  combination of renormalization group (RG) and self-consistent Hartree-Fock theories. 
Building on the foundational studies by Aleiner, Kharzeev, and Tsvelik \cite{aleiner2007spontaneous} and others \cite{kharitonov2012phase}, we use their RG scheme to integrate out states from the atomic-scale to the magnetic length $l_B$.  This approach specifically addresses the leading logarithmic divergences arising from effective interactions near the Dirac point, which are mediated by high-energy states close to the cutoff. We solve the RG flow equations using initial conditions derived from simple ``first-principles calculations'' \cite{wei2024landau}, leading to increases in Fermi velocity and complex adjustments to the short-range coupling constants. With these renormalized parameters in hand,  we then sum over another large class of Feynman diagrams using self-consistent Hartree-Fock calculations, evolving the density matrix towards a global minimum solution. The most surprising and remarkable aspect of this complicated approach is that it yields the $\nu-0$ phase diagram that is qualitatively consistent with experimental observations, achieved without any fine-tuning. This prompts us to apply the same method to predict the phase diagram at $\nu=\pm1$ in the absence of the boron-nitride potential.
This two-step approach, involving the use of perturbative renormalization group flow to account for quantum fluctuations followed by a non-perturbative self-consistent mean-field calculation, is akin in spirit to many earlier studies, Ref.~\cite{vafek2020renormalization,raghu2010superconductivity,kang2019strong}.

The remainder of this article is organized as follows. 
We begin with an overview of our main result, namely the $\nu=0$ and $\nu=\pm1$ phase diagram in Section.~\ref{sec:v=0 phase diagram}. In section.~\ref{sec:Model and RG}, we discuss the so-called standard model \cite{aleiner2007spontaneous,wei2024landau} of graphene, which is an effective theory characterized by nine parameters. Following this, we revisit the renormalization group calculations from Ref.~\cite{aleiner2007spontaneous}, providing a detailed analysis of each diagram, including the partial cancellation between double-exchange and BCS diagrams, and the repulsive nature of the vertex correction diagram. We then discuss the self-consistent Hartree-Fock calculations in Section.~\ref{sec:HF section}. We found that the converged density matrix exhibits a simple structure; while particle and hole Landau levels mix, such mixing occurs only within the same Landau level index in an isotropic Dirac fluid. We elaborate on the matrix elements of this converged density matrix and highlight the important role of the Dirac sea in contributing to the magnetic anisotropic energy. The article concludes with a summary and an outlook on future research in section.~\ref{sec:conclusion}.

\section{$\nu=0$ and $\nu=\pm1$ phase diagrams}\label{sec:v=0 phase diagram}

In this section, we discuss the main outcome of our work -- the $\kappa_0-B$ phase diagram shown in Fig.~\ref{fig:zero_LL_phase_diagram}. Previously, phase diagrams for generalized spin-valley quantum Hall ferromagnets in monolayer graphene (e.g.~those proposed by Kharitonov \cite{kharitonov2012phase}) were plotted against two coupling constants, $(u_\perp, u_z)$. These constants quantify the strength of the  valley-antisymmetric delta-function and inter-valley delta-function interactions projected onto the zeroth Landau level manifold.  As we noted in the introduction, the validity of using a zeroth-Landau-level-projected Hamiltonian to accurately infer the ground state order parameter is questionable. Indeed, Ref.~\cite{das2022coexistence,de2023global} showed that projecting general valley-dependent interactions onto the zeroth Landau level yield four parameters, not two. Nonetheless, these parameters have very complicated relationships to experimental variables, such as dielectric screening and magnetic field strength.
Using the two-step approach, which combines RG techniques with self-consistent Hartree-Fock calculations, we are able to trace how the coupling constants in graphene depend on experimental variables. This methodology enables us to express the phase diagram more directly in terms of $\kappa_0$  and $B$.

In Fig.~\ref{fig:zero_LL_phase_diagram}a represents the limiting case where the sublattice imbalance potential generated by the hexagonal Boron-Nitride (BN) substrate is zero ($\Delta_{BN}=0$), and Fig.~\ref{fig:zero_LL_phase_diagram}b represents the case where $\Delta_{BN}=1$meV.  The key takeaway from these phase diagrams is that as we move towards larger $\kappa_0$ and smaller $B$ -- the bottom right corner of the phase diagrams -- the renormalization effects from remote Landau levels become more important, and this prompts a phase transition from CAF to KD state. 
The two vertical dashed lines in this phase diagram correspond to the fine-structure constants for a double encapsulated sample ($\kappa_0\approx0.5$, $\epsilon_r\approx4.4$) and an open-surface sample ($\kappa_0\approx0.8$, $\epsilon_r\approx2.7$). 
In double-encapsulated samples typically used in transport experiments, a substantial portion of the phase space is dominated by the CAF phase, whereas in open-surface samples, typically used in STM experiments, most of the phase space is dominated by the KD. We note that the $\kappa_0$-induced transition was predicted by perturbation theory \cite{wei2024landau}, but the $B$-induced transition requires the RG machinery to resum the leading logarithmic divergences.
Our results suggest that the phase transition line is first order, although this conclusion is tentative. This is partly because when using RG to compute the 4-point vertex function, the standard approach \cite{shankar2017quantum} usually neglect the momentum dependence, considering them irrelevant in the RG sense. However, these ostensibly negligible terms could potentially alter the nature of the transitions from first to second order\cite{PhysRevLett.128.106803}.

\begin{figure}[t]
    \centering
    \includegraphics[width=0.95\linewidth]{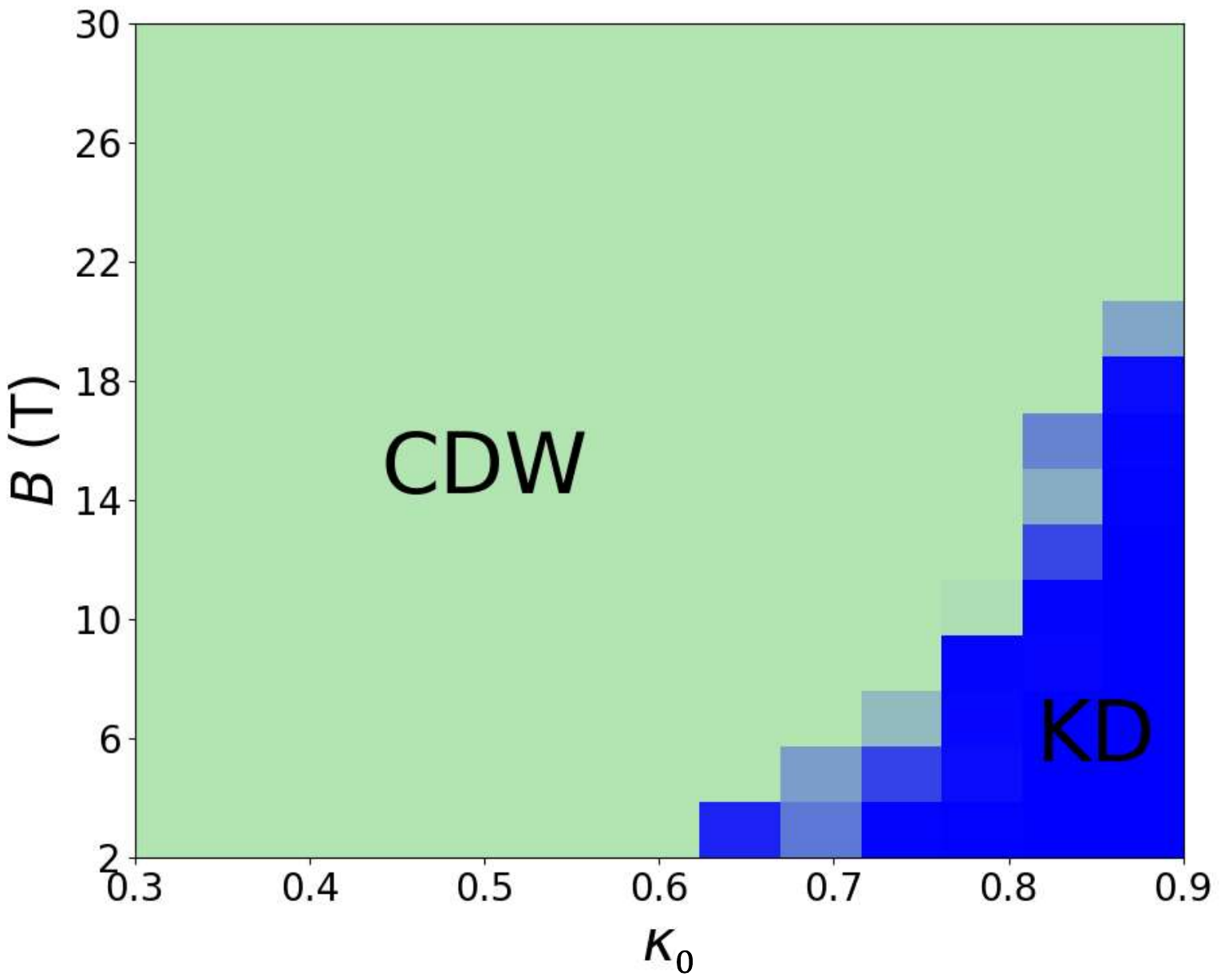}
    \caption{
        Zero-temperature phase diagram of       $ \nu = \pm 1 $ graphene computed with renormalized Fermi velocity and renormalized short-range interactions ($ \Delta_{BN} = 0$), incorporating both zeroth-Landau level and Dirac sea energy contributions. It shows a first-order phase transition between the spin-polarized charge density wave (CDW) and spin-polarized Kekul\'e-Distorted (KD) states.
    }
    \label{fig:phase-diagram-nu=1}
\end{figure}

Let us now summarize the phase diagram at $\nu=\pm1$. At this filling fraction, the system behaves as a well-known quantum Hall ferromagnet, where strong exchange interactions align the 4-component spin-valley pseudospins of all electrons (holes) in the same directions.  The relatively weak Zeeman field then aligns the collective spin with the total applied magnetic field. However, the fate of valley (sublattice) polarization remains uncertain when $\Delta_{BN}=0$. This uncertainty arises from the vanishing expectation value of the standard delta-function-type sublattice-valley dependent interaction \cite{kharitonov2012phase}, due to the cancellation between Hartree and Fock potentials in the zeroth-Landau-level-projected Hamiltonian at $\nu=\pm1$.
This Hartree-Fock cancellation is reminiscent of the well-known effect in the Hubbard model, where the spin-up self-energy receives repulsive contributions only from spin-down electron density, as contributions from spin-up density are exactly canceled due to the Pauli exclusion principle.

However, this vanishing expectation value of the spin-valley dependent interaction is an artifact of projecting the Hamiltonian onto the zeroth Landau level. When the projection extends beyond the zeroth Landau level to include a significant number of Landau levels in the mean-field calculations, we find that the Dirac sea Landau levels contribute to the self-energy of the zeroth Landau level in such a way that its Hartree and Fock potentials no longer cancel perfectly. This leads to the $\kappa_0-B$ phase diagram  shown in Fig.~2. We found that the Dirac sea favors spin-polarized charge-density waves in most of the $\kappa_0-B$ phase space and only become spin-polarized KD state at strong coupling. As we explained later, the driving force of this transition is due to the renormalization group flow of the coupling constants. 

%


\section{Model and Renormalization Group study}\label{sec:Model and RG}


In this section, we discuss the ``standard model'' of graphene and explore the renormalization of the coupling constants in the presence of the Coulomb interaction.  While much of the foundational work has been established in previous studies \cite{aleiner2007spontaneous,Son2007QuantumCP, Foster2008GrapheneVL,Gonzlez1998MarginalFermiliquidBF,GONZALEZ1994595,PhysRevLett.113.105502,PhysRevB.89.235431}, our goal is to provide a self-contained discussion and highlight key insights.  

Let's consider the following Euclidean action:
\begin{align}
&S_E = S_0 + S_b + S_{int}, \label{eq:total action} \\ 
&S_0[\psi^\dagger,\psi] = \int d^2r\,d\tau\,\sum_{s=\uparrow,\downarrow}\left(\psi^\dagger_s
    \partial_\tau\psi_s +\mathcal{H}_0(\psi^\dagger,\psi) \right),  \label{eq:fermion term} \\ 
    &\mathcal{H}_0(\psi^\dagger,\psi)=- iv_F\psi^\dagger_s \vec{\sigma} \cdot \nabla \psi_s \label{eq:H_0}\\
&S_b[\phi] = \int d^2r\, dz\, d\tau \sum_{i=x,y,z}\frac{1}{2}(\partial_i \phi)^2, \label{eq:boson term} \\ 
&S_{int}[\psi^\dagger,\psi,\phi] =i\sum_{s=\uparrow,\downarrow}\int d^2r\, d\tau\, g\psi^\dagger_s \psi_s \phi(r,0,\tau). \label{eq:interaction term}\\
&+\frac{1}{2}\sum^{\text{exclude $u=v=0$}}_{u,v=0,x,y,z}g_{uv}\int d\tau\int d^2 r \left(\psi_s^\dagger(\tau^{u} \sigma^v)\psi_s\right)^2   \label{eq:short range terms} 
\end{align}
Here $r$ is the in-plane position vector and $\tau$ is the imaginary time.  The basis of the four component valley-sublattice spinor $\psi_s$ is chosen such that the kinetic Hamiltionian $\mathcal{H}_0$ is the same in both valleys:
\begin{align} \label{eq:basis}
    \psi_s & =(\psi_{KA},\psi_{KB},\psi_{K^\prime B},-\psi_{K^\prime A})^{T},\\
    \psi &= (\psi_{\uparrow},\psi_{\downarrow})^{T}.
\end{align}
Here $\tau^\alpha$ are the Pauli matrices in valley space and $\sigma^\beta$ are the Pauli matrices in $\bar{A}\bar{B}$ space. 

Eq.~\eqref{eq:H_0} is the massless Dirac equation for a four-component fermion where \(v_F =10^6 \text{ m/s}\) is the Fermi velocity. The linear dispersion is accurate up to an energy cutoff of approximately \(\Lambda \sim 2 \text{ eV}\).

In Eq.~\eqref{eq:boson term} and Eq.~\eqref{eq:interaction term},  we introduced an auxiliary boson field \( \phi(r, z, \tau) \) via a Hubbard-Stratonovich transformation to describe the long-range Coulomb potential. The coupling-constant between the boson and the relativistic fermion is given by the following:
\begin{align}
    g = \frac{e}{\sqrt{\epsilon_0 \epsilon_r}} ,
\end{align}
where \(\epsilon_0\) is the vacuum permittivity, and \(\epsilon_r\) is the device dependent dielectric screening constant of the environment.  One can integrate out the boson field \(\phi\) in the path integral to recover the usual four-term long-range Coulomb interaction:
\begin{align}\label{eq:four terms coulomb interaction}
    H_{e-e} =\frac{1}{2}\int d^2 r d^2 r^\prime \psi^\dagger(r)\psi^\dagger(r^\prime) \frac{e^2}{4\pi \epsilon_0 \epsilon_r|r-r^\prime|}\psi(r^\prime)\psi(r)
\end{align}
Eq.~\eqref{eq:boson term} represents the kinetic term for the boson field; however, it lacks a time derivative. This is because we model the long-range Coulomb interaction Eq.~\eqref{eq:four terms coulomb interaction} as an instantaneous interaction. In this representations, the boson lives in three-dimensional space while the electrons are confined to the 2D  $z=0$ plane of the material.

\begin{figure}[h]
    \centering
    \includegraphics[width=1\linewidth]{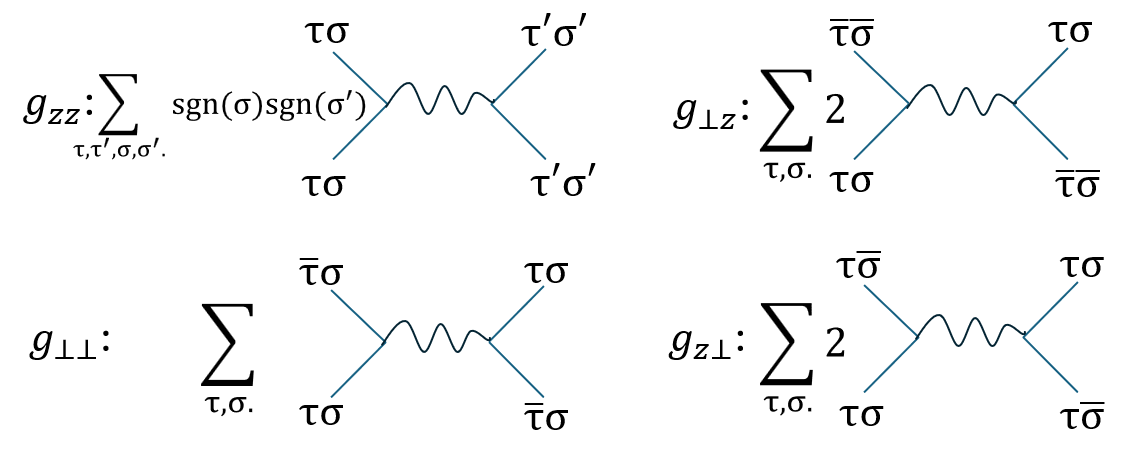}
    \caption{This figure provides a diagrammatic representation of the four valley dependent anisotropic coupling constants arising from bare electron-electron interactions. The solid lines denote the incoming and outgoing electrons, respectively, while the curved line represents the Coulomb interaction. Here, \(\bar{\tau}\) and \(\bar{\sigma}\) denote opposite valleys and sublattices, respectively. The sign convention is chosen such that \( \text{sgn}(A) = \text{sgn}(K) = 1 \). From the diagram, we observe that for \( g_{zz} \), neither the valley nor the sublattice indices change; however, the sign changes when the interaction occurs at different sublattices. In contrast, \( g_{\perp z} \) induces a change in both the valley and sublattice indices. For \( g_{\perp \perp} \), only the valley index changes while the sublattice index remains unchanged. Conversely, \( g_{z\perp} \) alters the sublattice index while preserving the valley index.
}
    \label{fig:short-range coupling diagram}
\end{figure}

Eq.~\eqref{eq:short range terms} describes the spin-valley dependent short-range interaction on the lattice scale, with $ g_{uv} $ denoting their coupling constants. Although there are generally fifteen different terms, the short-range coupling constants take the following form due to the \( C_{6v} \) and translational symmetries \cite{wolf2024magnetismdiluteelectrongas}:

\begin{align}
    g_{z\perp} & = g_{zx} = g_{zy}, \\
    g_{\perp z} & = g_{xz} = g_{yz}, \\
    g_{\perp \perp} & = g_{xx} = g_{xy} = g_{yx} = g_{yy}, \\
    g_{0\perp} & = g_{0x} = g_{0y}, \\
    g_{\perp0} & = g_{x0} = g_{y0},
\end{align}
Additionally, due to time-reversal symmetry, the couplings \( g_{z0} \), \( g_{0z} \), \( g_{\perp0} \), and \( g_{0\perp} \) all vanish. Consequently, only four independent coupling constants remain: \( g_{zz} \), \( g_{\perp z} \), \( g_{z\perp} \), and \( g_{\perp \perp} \). The physical significance of these four coupling constants is illustrated in Fig.~\eqref{fig:short-range coupling diagram}.

\begin{figure*}
    \centering
    \includegraphics[width=0.9\linewidth]{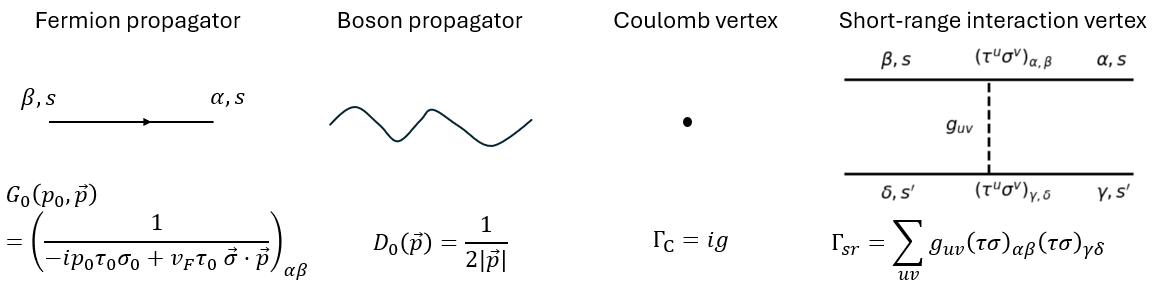}
    \caption{The Feynman rules in momentum space, where \(\vec{p}\) represents the momentum in the sample plane and \(p_0\) denotes the Matsubara frequency. The last term is the diagrammatic representation of the short-range interaction. The indices \(\alpha\), \(\beta\), \(\gamma\), and \(\delta\) denote the valley-sublattice indices, while \(s\) represents the spin index.}
    \label{fig:Feynman rules}
\end{figure*}

In Fig.~\eqref{fig:Feynman rules}, we present the Feynman diagrams along with the corresponding Feynman rules for the action, as given by Eq.~\eqref{eq:total action},  in momentum space. In the following section, for simplicity, we will denote \( \tau^0\sigma^i \) simply as \( \sigma^i \), \(\tau^i \sigma^0\) as \(\tau^i\) and $\tau^0\sigma^0$ as $I$ without further explanation.

\subsection{RG flow of Fermi velocity and quasiparticle-residue}
In this section, we derive the RPA self-energy of the electron and resum its leading log-divergence using standard RG approach. The RPA electron self-energy is shown in Fig.~\eqref{fig:electron self energy}a) where the double dashed line denotes the dynamically screened Coulomb interaction within RPA.

We begin by computing the screened Coulomb interaction using RPA. In RPA, the screening effect is modeled as a geometric series of the particle-hole susceptibility, $\Pi$, and the bare (direct) Coulomb interaction, $V_q$. The particle-hole susceptibility is computed using its standard expression:

\begin{figure}[h]
    \centering
    \includegraphics[width=1\linewidth]{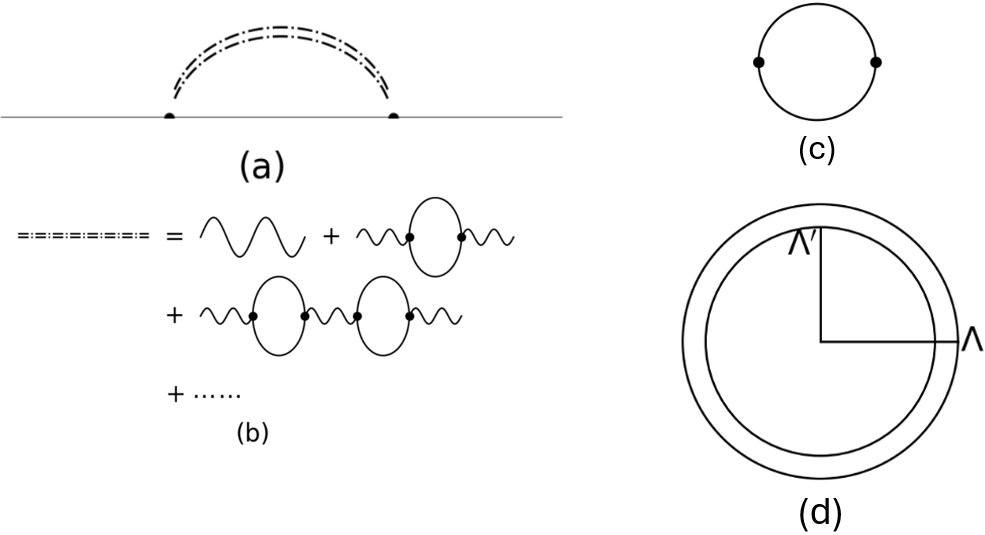}
    \caption{a) Electron self-energy generated by the screened Coulomb interaction within random phase approximation (RPA). b) RPA-modified Coulomb interaction. c) Particle-hole susceptibility. d) The states being integrated out in the RG procedure live in a thin-shell of $p_0,|\vec{p}|$ space: \( |\Lambda^\prime|^2 < p_0^2 + v_F^2|\vec{p}|^2 < |\Lambda|^2 \).
}
    \label{fig:electron self energy}
\end{figure}

\begin{align}
    \Pi(\vec{p}, p_0) &=-N\int \frac{d^3q}{(2\pi)^3}\text{Tr}\left(G_0(p_0+q_0,\vec{p}+\vec{q}) G_0(q_0,\vec{q}) \right)\\
    &=\frac{N}{8} \frac{|\vec{p}|^2}{\sqrt{p_0^2 + v_F^2 |\vec{p}|^2}} \label{eq:vacuum polarization}
\end{align}
Here \(p_0\) Matsubara frequency and $\vec{p}$ is the momentum. Here $G_0$ is a 4 by 4 matrix in valley and sublattice space and $N = 2$ represent spin degeneracy. From this, we evaluate the dressed boson propagator $\Bar{D}(\vec{p}, p_0)$ [Fig.~\eqref{fig:electron self energy}b)] that represents the dynamically screened Coulomb potential as a three-point vertex function:
\begin{align}\label{eq:dressed boson propagator}
    \Bar{D}(\vec{p},p_0) & = \frac{D_0(\vec{p})}{1+g^2D_0(\vec{p})\Pi(\vec{p},p_0)}\\
    & = \frac{1}{2|\vec{p}|+\frac{Ng^2}{8}\frac{|\vec{p}|^2}{\sqrt{p_0^2+v_F^2|\vec{p}|^2}}}
\end{align}

%
 Next, we calculate the electron self-energy resulting from scattering off these bosons. Following the Wilsonian RG, we focus on the contributions from very short-wavelength and high-frequency modes. These modes reside within a thin shell defined between $\Lambda$ and $\Lambda^\prime$, as shown Fig.~\eqref{fig:electron self energy}d,

\begin{align}
\Sigma_C&= \int^{\Lambda}_{\Lambda^\prime} \frac{d^3q}{(2\pi)^3} ig G_0(p_0-q_0,\vec{p}-\vec{q}) ig \bar{D}(\vec{q},q_0)  \label{eq:screened self energy} \\
& = \int^{\Lambda}_{\Lambda^\prime} \frac{d^3q}{(2\pi)^3} ig \frac{i(p_0-q_0)+v_F\sigma\cdot(\vec{p}-\vec{q})}{(p_0-q_0)^2+v_F^2(\vec{p}-\vec{q})^2} ig \bar{D}(\vec{q},q_0)\\
& = -g^2\int^{\Lambda}_{\Lambda^\prime} \frac{d^3q}{(2\pi)^3} \frac{i(p_0-q_0)+v_F\sigma\cdot(\vec{p}-\vec{q})}{q_0^2+v_F^2\vec{q}^2} \\
&\times(1-2\frac{p_0q_0+v_F^2\vec{p}\cdot\vec{q}}{q_0^2+v_F^2\vec{q}^2}+\frac{p_0^2+v_F^2\vec{p}^2}{q_0^2+v_F^2\vec{q}^2})^{-1} \bar{D}(\vec{q},q_0)  \notag \\
& \approx -g^2\int^{\Lambda}_{\Lambda^\prime} \frac{d^3q}{(2\pi)^3} \frac{i(p_0-q_0)+v_F\sigma\cdot(\vec{p}-\vec{q})}{q_0^2+v_F^2\vec{q}^2}\label{eq:all terms} \\
&\times(1+2\frac{p_0q_0+v_F^2\vec{p}\cdot\vec{q}}{q_0^2+v_F^2\vec{q}^2})\bar{D}(\vec{q},q_0) \notag\\
& =  -g^2\int^{\Lambda}_{\Lambda^\prime} \frac{d^3q}{(2\pi)^3} \frac{i(v_F^2\vec{q}^2-q_0^2)p_0+v_F^2q_0^2 \vec{\sigma}\cdot\vec{p}}{(q_0^2+v_F^2\vec{q}^2)^2}\bar{D}(\vec{q},q_0)\label{eq: odd terms} \\
& = f_1 ip_0-f_2v_F \vec{\sigma}\cdot\vec{p}
\end{align}

where we define $f_1$ and $f_2$ as:
\begin{align}
f_1& =g^2\int^{\Lambda}_{\Lambda^\prime} \frac{d^3q}{(2\pi)^3} \frac{(q_0^2-v_F^2\vec{q}^2)}{(q_0^2+v_F^2\vec{q}^2)^2}\bar{D}(\vec{q},q_0)\\ 
& = \frac{Ng^2}{2v_F}\int\frac{d\theta}{(2\pi)^2}\frac{\cos{2\theta}}{1+\lambda \sin{\theta}}\int_{\Lambda^\prime}^{\Lambda}\frac{dq}{q}\\
& =  -\frac{8}{N\pi^2}\left(  1-\frac{\pi}{2\lambda}+\frac{(2-\lambda^2)\arccos(\lambda)}{2\lambda\sqrt{
    1-\lambda^2}}   \right)\ln{\left(  \frac{\Lambda}{\Lambda^\prime}   \right)}  \label{eq:define f1} 
\end{align}
\begin{align}
    f_2 &=g^2\int^{\Lambda}_{\Lambda^\prime} \frac{d^3q}{(2\pi)^3} \frac{q_0^2}{(q_0^2+v_F^2\vec{q}^2)^2}\bar{D}(\vec{q},q_0)\\
    &=\frac{Ng^2}{2v_F}\int\frac{d\theta}{(2\pi)^2}\frac{\cos^2{\theta}}{1+\lambda \sin{\theta}}\int_{\Lambda^\prime}^{\Lambda}\frac{dq}{q}\\
    & =-\frac{4}{N\pi^2}\left(  1-\frac{\pi}{2\lambda}+\frac{\sqrt{1-\lambda^2}}{\lambda}\arccos(\lambda)\right) \ln{\left(  \frac{\Lambda}{\Lambda^\prime}   \right). }\label{eq:define f2}
\end{align}
Here  $\lambda$ is a dimensionless parameter proportional to the fine structure constant:
\begin{equation} \label{eq:lambda_def}
    \lambda =\frac{Ng^2}{16v_F} = \frac{N\pi}{4} \kappa
\end{equation}
Here $\lambda\sim3.4$ when $\kappa=2.19$ ($\epsilon_r=1$). We emphasis that $\kappa$ here is defined with renormalized Fermi velocity, and should be distinguished from $\kappa_0$, which is defined using the bare Fermi velocity.
From Eq.~\eqref{eq:all terms} to Eq.~\eqref{eq: odd terms}, we retain only the  terms even in \(q_0\) and \(\vec{q}\), as all odd terms vanish upon integration. Additionally, we replace \((\vec{\sigma}\cdot\vec{q})(\vec{p}\cdot\vec{q})\) with \(\frac{1}{2}\vec{q}^2(\vec{\sigma}\cdot\vec{p})\) due to the isotropy of the integration. These simplifications extract the leading logarithmic dependence on the energy cutoffs.

Using the self-energy expression, we derive the following expression for the dressed Green function:
\begin{align}
G(p_0,\vec{p},\Lambda^\prime,\Lambda) & = \frac{G_0}{1-G_0\Sigma_C}\\
& = \frac{1}{-i(1+f_1)p_0+v_F(1+f_2)\vec{\sigma}\cdot\vec{p}}.
\end{align}
where $G_0$ is the bare Green function.

To derive the RG flow equation, we consider the renormalized Green function:
\begin{align}
G_{\text{r}}(p_0,\vec{p},  v_F') &\equiv Z^{-1} G(p_0,\vec{p}, v_F, \Lambda^\prime)\\
& = \frac{1}{-ip_0+v_F^\prime \vec{\sigma}\cdot\vec{p}}
\end{align}
 where $v_F'$ represents the renormalized Fermi velocity. The factor $Z$, known as the quasiparticle residue, quantifies the weight of the quasiparticle peak in the spectral function. By fixing the coefficient of $p_0$ in $G_{\text{r}}$ to 1, we obtain,

\begin{align}
    Z(\Lambda^\prime) &= \frac{1}{1+f_1}\\ 
    v_F^\prime(\Lambda^\prime) & = \frac{1+f_2}{1+f_1}v_F(\Lambda)   .\label{eq:fermi velocity}
\end{align}
We then proceed to reduce set $\Lambda^\prime = \Lambda - d\Lambda$ and reproduce the following flow equations
\cite{Son2007QuantumCP, Gonzlez1998MarginalFermiliquidBF,PhysRevLett.113.105502}:
\begin{align}
    \frac{d v_F(\Lambda)}{d \ln{\Lambda}} &= -\frac{4}{\pi^2 N}\left(1-\frac{\pi}{2\lambda}+\frac{\arccos(\lambda)}{\lambda\sqrt{1-\lambda^2}}    \right) v_F(\Lambda) \label{eq:v_F2} \\
    \frac{dZ(\Lambda)}{d\ln{\Lambda}}
    & = \frac{8}{N\pi^2} \left( 1 - \frac{\pi}{2\lambda} + \frac{(2 - \lambda^2)\arccos(\lambda)}{2\lambda\sqrt{1 - \lambda^2}} \right) Z(\Lambda) \label{eq:RG Z}
\end{align}
These expressions describe how the Fermi velocity and quasiparticle residue changes as we systematically integrate out short-distance, high energy fluctuations. Since the electric charge does not change upon renormalization, we can write Eq.~\eqref{eq:v_F2} in terms of the dimensionless parameter 
$\lambda$, (Eq.~\eqref{eq:lambda_def}) which is proportional to the fine-structure constant:
\begin{align} \label{eq:RG Lambda}
    \frac{d \lambda(\Lambda)}{d \ln{\Lambda}} &= \frac{4}{\pi^2 N}\left(1-\frac{\pi}{2\lambda}+\frac{\arccos(\lambda)}{\lambda\sqrt{1-\lambda^2}}    \right) \lambda(\Lambda)
\end{align}



\begin{figure}[h]
    \centering
    \includegraphics[width=0.8\linewidth]{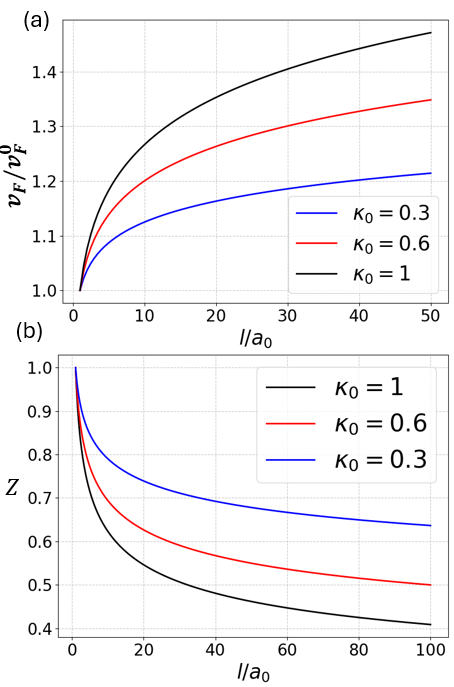}
    \caption{a) shows the renormalization of the Fermi velocity $v_F(l)$ at different values of $\kappa_0$. $l\sim1/\Lambda$. 
    Here, \( a_0 = 2.5\,\text{\AA} \) represents the graphene lattice constant. b) shows renormalization of the quasiparticle residue Z(l). The reduction in \( Z \) becomes more pronounced at higher \(\kappa_0\). However, even at sufficiently large length scales (corresponding to low energy scales), the quasiparticle residue \( Z \) does not decrease to zero.
 }
    \label{fig:Fermi velocity}
\end{figure}

Let us now solve for Eq.~\eqref{eq:RG Lambda} and Eq.\eqref{eq:RG Z}.
Firstly, we numerically integrate Eq.~\eqref{eq:RG Lambda} [or equivalently Eq.~\eqref{eq:v_F2}] with an initial Fermi velocity $v_F^0=1\times 10^6\text{m/s}$. This process yields the renormalized Fermi velocity as a function of RG time, defined as $l/a$ where
$a$ is graphene's lattice constant and 
$l$ is a larger length scale of our effective long-wavelength theory. In the presence of a magnetic field, we can set the cutoff to the magnetic length. Once we have $v_F$ vs $l/a$, we can then 
 integrate Eq.~\eqref{eq:RG Z} and obtain $Z$ as a function of RG time. The results are shown in Fig.~\eqref{fig:Fermi velocity}.   Fig.~\eqref{fig:Fermi velocity}a) shows that the Fermi velocity increases from its bare value $v_F^0$ as we move towards the infrared (IR) scale, with larger $l/a$ ratios. For small fine structure constant $\kappa_0$, i.e.~large dielectric screening, this enhancement is mainly coming from the first order exchange effect. For large $\kappa_0$, i.e.~weak dielectric constant,  the enhancement is more pronounced due to contributions from higher-order diagrams. In Fig.~\eqref{fig:Fermi velocity}(b), the quasiparticle residue $Z$ is shown to decrease monotonically as we approach the IR scale. However, the rate of change diminishes as $l$ approaches infinity, leading to the concept of a marginal Fermi liquid as suggested by Ref.~\cite{Gonzlez1998MarginalFermiliquidBF}.

\subsection{RG equation of the sublattice-valley dependent coupling constants}

In this subsection, we study the RG flow of sublattice-valley dependent coupling constants. While these coupling constants might appear weak at the atomic-scale, their magnitude can grow and sign might change upon RG by mixing with the strong screening Coulomb interaction. This will then influence the magnetic anisotropic energy landscape in the zeroth Landau level.  Indeed, ab initio calculations \cite{wei2024landau} using a linear combination of 2$p_z$ atomic orbitals indicate that the computed values of these coupling constants are significantly smaller than those needed to explain experiments \cite{zibrov2018even}, suggesting these constants receive a strong many-body corrections.

These sublattice-valley depedendent couplings receive contributions from both electron-electron (e-e) (c.f.~Fig.~\eqref{fig:short-range coupling diagram})) and electron-phonon (e-p) interactions. We designate the contribution from e-e interaction as  $g_{\alpha\beta}^{e-e}$
  and from e-p interaction as  $g_{\alpha\beta}^{e-p}$. The total coupling constant is $ g_{\alpha\beta} = g_{\alpha\beta}^{e-p}+g_{\alpha\beta}^{e-e}$.
 The electron-electron and electron-phonon interactions will be analyzed separately.
 The diagrammatic representation of the short range interaction is given by Fig.~\eqref{fig:Feynman rules}.



\begin{figure*}
    \centering
    \includegraphics[width=1\linewidth]{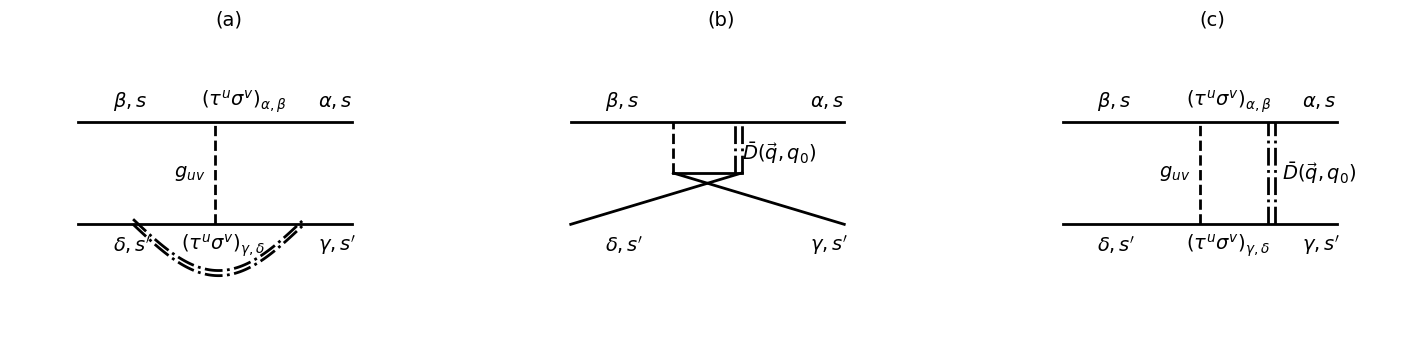}
    \caption{Feynman diagrams consists of one short-range interaction and one long range interaction. Here a) is the vertex correction diagram, b) is the double exchange diagram, and c) is the BCS diagram.  The external momenta are set to zero.
}
    \label{fig:short range diagram}
\end{figure*}

We begin with the e-e interaction and consider the mixed one-loop diagram where one vertex is delta-function short-range interactions and the other vertex is the dynamically screened long-range Coulomb interaction with RPA.
At this one-loop level, the four point vertex functions receive four distinct contributions from the Bardeen-Cooper-Schrieffer (BCS) diagram, the double-exchange (DE) diagram, the vertex corrections (VC) diagrams, and those associated with wavefunction renormalization ($Z$). We note that the bubble diagram vanishes, as explained in Ref.~\cite{wei2024landau}. The DE and VC diagrams are generated by particle-hole propagation, whereas the BCS diagram arises from particle-particle propagation. In semimetals like neutral graphene where the number of holes (occupied states) and particles (unoccupied states) are the same, no single diagram dominates the interaction processes and we have to account for all of them:
\begin{equation}
    \Gamma = \Gamma^{BCS} + \Gamma^{DE} + \Gamma^{vc} + \Gamma^{Z}.
\end{equation}
We note that these 4-point vertex functions are spin-independent, 
\begin{equation}
    \Gamma_{\alpha s_1 \beta s_2,\gamma s_3\delta s_4}=\Gamma_{\alpha \beta, \gamma  \delta }\,\delta_{s_1,s_3}\delta_{s_2,s_4}.
\end{equation}
The evaluation of the four-point vertex function is very similar to the computation of the self-energy. It turns out that the logarithmically divergent part of the BCS diagram can be expressed in terms of functions $f_1$
and $f_2$ that also appear in the self-energy calculations:
\begin{align} \label{eq:Gamma_BCS}
    &\Gamma^{BCS}_{\alpha \beta ,\gamma  \delta } \notag \\
     &= -\sum_{\substack{u = x, y, z \\ \perp = x, y}}\left[g_{uz}^{e-e}f_2+g_{u\perp}^{e-e} \frac{f_2-f_1}{2}\right](\tau^u \sigma^z)_{\alpha,\beta}(\tau^u \sigma^z)_{\gamma,\delta} \notag \\
    &- \sum_{\substack{u = x, y, z \\ \perp = x, y}}\left[g_{uz}^{e-e}\frac{f_2-f_1}{2}+g_{u\perp}^{e-e}f_2 \right](\tau^{u} \sigma^\perp)_{\alpha,\beta}(\tau^{u} \sigma^\perp)_{\gamma,\delta}\notag \\ 
    &- \sum_{\substack{u = x, y, z \\ \perp = x, y}}\left[ g_{u\perp}^{e-e} \frac{f_2-f_1}{2}\right](\tau^{u} i \sigma^0)_{\alpha,\beta}(\tau^{u} i \sigma^0)_{\gamma,\delta}.
\end{align}
Here $f_1$ and $f_2$ are defined in Eq.~\eqref{eq:define f1} and Eq.~\eqref{eq:define f2} with $f_2>f_1>0$.  Since electronic contributions to the coupling constants are repulsive $g^{e-e}>0$, the BCS diagram generates an attractive interaction that reduces the bare interaction strength.
We recall that the first and second indices of $g_{u\alpha}$ correspond to the valley and sublattice degrees of freedom, respectively, in the basis of Eq.~\eqref{eq:basis}. 
As the total valley number is a conserved quantity, the intra-valley and inter-valley renormalization processes happens independently. However, because the $\pi$-band eigenstate is a superposition of graphene's sublattices, intra-sublattice and inter-sublattice interactions are mixed in the two-particle propagation process. Thus, the first square bracket in  Eq.~\eqref{eq:Gamma_BCS}, describing the interaction generated in the $\tau^u\sigma^z$ channel, receives contributions from both $g_{uz}$ and $g_{u\perp}$. Similarly, 
the second square bracket in Eq.~\eqref{eq:Gamma_BCS}, describing the interaction generated in the $\tau^u\sigma^\perp$ channel, also involves contributions from both types. The last term introduces a new coupling constant $g_{u0}$ not present at the bare level.


The next diagram we discussed is the double-exchange  (DE) diagram, it is given by the following:
\begin{align}
    &\Gamma^{DE}_{\alpha \beta ,\gamma \delta}   \notag \\
     &=\sum_{\substack{u = x, y, z \\ \perp = x, y}} \left[ g_{uz}^{e-e}f_2-g_{u\perp}^{e-e} \frac{f_2-f_1}{2} \right] (\tau^u \sigma^z)_{\alpha,\beta}(\tau^u \sigma^z)_{\gamma,\delta} \notag \\
    &- \sum_{\substack{u = x, y, z \\ \perp = x, y}} \left[ g_{uz}^{e-e}\frac{f_2-f_1}{2}-g_{u\perp}^{e-e}f_2 \right](\tau^{u} \sigma^\perp)_{\alpha,\beta}(\tau^{u} \sigma^\perp)_{\gamma,\delta} \notag \\ 
    &+ \sum_{\substack{u = x, y, z \\ \perp = x, y}} 
    \left[ g_{u\perp}^{e-e} \frac{f_2-f_1}{2} \right] (\tau^{u} i \sigma^0)_{\alpha,\beta}(\tau^{u} i \sigma^0)_{\gamma,\delta}.
\end{align}
Similar to the BCS diagram, the first term correspond to renormalization in the $\tau^u\sigma^z$ channel, the second term corresponds to renormalization in the $\tau^u\sigma^\perp$ channel. It's important to note that the third term in the DE diagram is exactly canceled out by the corresponding term in the BCS diagram. Consequently, no new coupling constants are introduced at the one-loop level; we only observe the renormalization of the existing coupling constants that were present at the ``bare level''.

When comparing the $\tau^u\sigma^z$ channel from the DE diagram with those from the BCS diagram, we observe that the contribution from $g_{uz}^{e-e}$ cancels out, leaving only the contribution from $g_{u\perp}^{e-e}$. A similar cancellation happens in the $\tau^u\sigma^\perp$ channel. 

Next, the vertex-correction (VC) diagram is given by the following:
\begin{align}
   \Gamma^{vc}_{\alpha \beta, \gamma  \delta  } 
    & =\sum_{u=x,y,z} g_{uz}^{e-e}(2f_2-f_1) (\tau^u \sigma^z)_{\alpha,\beta}(\tau^u \sigma^z)_{\gamma,\delta} \notag \\
    & +\sum_{\substack{u = x, y, z \\ \perp = x, y}} g_{u\perp}^{e-e}f_2(\tau^{u} \sigma^\perp)_{\alpha,\beta}(\tau^{u} \sigma^\perp)_{\gamma,\delta}
\end{align}
Unlike the BCS and double exchange diagrams, the vertex correction diagram does not involve mixing of the sublattice degree of freedom. Instead, it enhances the repulsive interactions already present, making them more repulsive.  

In addition to the three one-loop diagrams, wave function renormalization must also be accounted for, as described in Eq.~\eqref{eq:short range terms}:
\begin{align}
    \Gamma^{Z}_{\alpha \beta ,\gamma  \delta}&= 2\delta Z \times \sum_{u,v=x,y,z} g_{uv}(\tau^{u} \sigma^{v})_{\alpha,\beta}(\tau^{u} \sigma^{v})_{\gamma,\delta} \\
    &= -2f_1\sum_{u=x,y,z} g_{uz}(\tau^{u} \sigma^z)_{\alpha,\beta}(\tau^{u} \sigma^z)_{\gamma,\delta} \notag \\
    & -2f_1\sum_{\substack{u = x, y, z \\ \perp = x, y}} g_{u\perp}(\tau^{u} \sigma^{\perp})_{\alpha,\beta}(\tau^{u} \sigma^{\perp})_{\gamma,\delta}
\end{align}

After summing over all the contributions -- including the three diagrams and the wave function renormalization -- we arrive at the following correction to the coupling constants at one-loop order:
\begin{align}
    &\delta g_{uz}^{e-e} =4(f_2-f_1)(g_{uz}^{e-e}-g_{u\perp}^{e-e}),\label{eq:one loop change in guz} \\ 
    & \delta g_{u\perp}^{e-e} = 2(f_2-f_1)(g_{u\perp}^{e-e}-g_{uz}^{e-e}). \label{eq:one loop change in gux}
\end{align}
This leads to the following RG equations \cite{aleiner2007spontaneous}:
\begin{align}
    &\frac{dg_{uz}^{e-e}}{d\ln{(\Lambda )}} = -\frac{16}{N\pi^2}\left(1-\frac{\pi}{2\lambda}+\frac{\arccos(\lambda)}{\lambda\sqrt{1-\lambda^2}}   \right)(g_{uz}^{e-e}-g_{u\perp}^{e-e})\\
      &\frac{dg_{u\perp}^{e-e}}{d\ln{(\Lambda )}} = -\frac{8}{N\pi^2}\left(1-\frac{\pi}{2\lambda}+\frac{\arccos(\lambda)}{\lambda\sqrt{1-\lambda^2}}   \right)(g_{u\perp}^{e-e}-g_{uz}^{e-e})
\end{align}
Let the RG time be $\Lambda=a/l$, where $l$ is a length scale of interest and $a$ is graphene's lattice constant. A straightforward integration of the RG equations yields the relationship between the renormalized coupling constants at length scale $l$, $g_{\alpha\beta}^{e-e}(l)$, and the bare coupling constants evaluated at the lattice constant $a$, $g_{\alpha\beta}^{e-e}(a)$:
\begin{widetext}
\begin{align}
    g_{uz}^{e-e}(l) &= \frac{1}{3}\left( g^{e-e}_{uz}(a) +2g^{e-e}_{u\perp}(a) \right) + \frac{2}{3}\left( g_{uz}^{e-e}(a)-g_{u\perp}^{e-e}(a) \right)e^{\int_{a}^{l}\frac{F(\lambda(l^\prime))}{l^\prime} dl^\prime}\label{eq:e-e contribution to guz} \\
        g_{u\perp}^{e-e}(l) &= \frac{1}{3}\left( g^{e-e}_{uz}(a) +2g^{e-e}_{u\perp}(a) \right) - \frac{1}{3}\left( g_{uz}^{e-e}(a)-g_{u\perp}^{e-e}(a) \right)e^{\int_{a}^{l}\frac{F(\lambda(l^\prime))}{l^\prime} dl^\prime}\label{eq:e-e contribution to gux}\\
        F(\lambda) &= \frac{24}{N\pi^2}\left(1-\frac{\pi}{2\lambda}+\frac{\arccos(\lambda)}{\lambda\sqrt{1-\lambda^2}}   \right)
\end{align}
\end{widetext}
Since $g^{e-e}_{z\perp} \approx g^{e-e}_{\perp z} \ll g^{e-e}_{zz}(a) < g^{e-e}_{\perp \perp}(a)$(see next subsection), we can infer from these equations that $g^{e-e}_{zz}(l)$ and $g^{e-e}_{\perp \perp}(l)$ become increasingly repulsive. In contrast, although $g^{e-e}_{z\perp}$ and $g^{e-e}_{\perp z}$ are initially positive, they become less positive as we approach the IR scale, eventually changing sign and becoming negative.

Next, we analysis the electron-phonon interactions. We focus on the dominant contributions from optical phonons. The interaction matrix elements for electron-optical phonon scattering changes the sublattice-projection of the electron, thus they only lead to renormalization in the coupling constants $ g_{z\perp} $ and $g_{\perp z}
$\cite{PhysRevB.77.041409,kharitonov2012phase}. The derivation of the electron-phonon RG equation is very similar to the electron-electron counterpart. Following the steps described earlier, we arrive at the following expression  \cite{basko2008interplay}:
\begin{align}
    \delta g_{\perp z}^{e-p} &=4(f_2-f_1)g_{\perp z}^{e-p} \\ 
     \delta g_{z\perp}^{e-p} &= 2(f_2-f_1)g_{z\perp}^{e-p}.
\end{align} 
Direct integration of RG time yields,

\begin{align}
    g_{z\perp}^{e-p}(l) &=g^{e-p}_{z\perp}(a) e^{\int_{a}^{l}\frac{F(\lambda(l^\prime))}{3l^\prime}dl^\prime} \label{eq:e-p contribution to gzx} \\
    g_{\perp z}^{e-p}(l) &=g^{e-p}_{\perp z}(a) e^{\int_{a}^{l}\frac{2F(\lambda(l^\prime))}{3l^\prime}dl^\prime}\label{eq:e-p contribution to gxz}
\end{align}
Since $g_{z\perp}^{e-p}$ and $g_{\perp z}^{e-p}$ are negative, their RG flow will drive these values even further negative, making $g_{z\perp}$ and $g_{\perp z}$ increasingly attractive.

\subsection{RG flow of short-Range coupling constants}

The bare values of $g_{uv}^{e-p}$ can be estimated by neglecting the retardation effects \cite{piscanec2004kohn,Wu2018TheoryOP}, with $g_{z\perp}^{e-p}=-52 meV\cdot nm^2$ and $g_{\perp z}^{e-p}=-69 meV\cdot nm^2$. Ref.~\cite{wei2024landau} provides a first-principle estimation of the bare values $g_{uv}^{e-e}$ by using the linear-combination of atomic orbitals assumptions, this leads to the values presented in Tab.~\eqref{tab:g value table}.
\begin{table}[h]
\centering
\begin{tabular}{c|c|c|c|c}
    $meV\cdot nm^2$ & \( g_{zz} \) & \( g_{\perp z} \) & \( g_{z\perp} \) & \( g_{\perp \perp} \) \\
    \hline
    e-e & 184 & 43 & 25 & 269 \\
    \hline
    e-p & 0 & -69 & -52 & 0 \\
    \hline
    Total & 184 & -26 & -27 & 269 \\
\end{tabular}
\caption{The bare values of the coupling constants for both the electron-electron and electron-phonon contributions.}
\label{tab:g value table}
\end{table}

\begin{figure}[h!]
    \centering
    \includegraphics[width=1\linewidth]{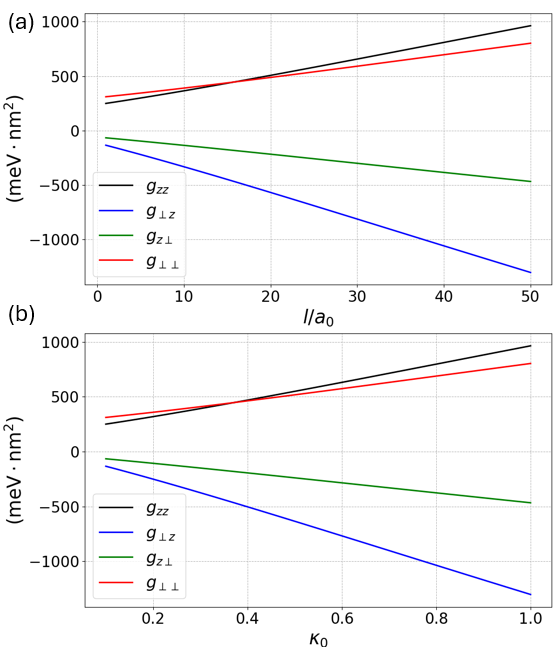}
    \caption{Flow of coupling constants vs RG time $l/a_0$ at a fixed fine structure constant \(\kappa_0 = 1\). Here $a_0$ is graphene lattice constant. b) Coupling constants vs $\kappa_0$ at the cutoff $l = l_B$, where the magnetic field used is $B = 10\,\text{T}$.
}
    \label{fig:RG flow of coupling}
\end{figure}

We will use the above bare values $g_{uz}(a_0)$ as initial values to solve the RG differential equation. The electron-phonon channel ($g_{\alpha\beta}^{e-p}(l)$) and electron-electron channel ($g_{\alpha\beta}^{e-e}(l)$) are renormalized independently and these results are added at the final stage.  The final result is shown in Fig.~\ref{fig:RG flow of coupling}.

Fig.~\ref{fig:RG flow of coupling}, shows that the magnitude of all coupling constants are enhanced with RG time ($l/a_0$). The coupling constants that is sublattice off-diagonal, namely $g_{z\perp}$ and $g_{\perp z}$ become more negative, while the sublattice diagonal ones, $g_{zz}$ and $g_{\perp \perp}$ become more positive. 
We note that $g_{\perp z}$ (blue-line in Fig.~\ref{fig:RG flow of coupling}) eventually becomes the most dominant coupling constant (largest in amplitude) at long wavelength limit. For magnetic field strengths such as $B=10T$, the RG flow should stop at $l_B/a\sim40$. Fig.~\eqref{fig:RG flow of coupling}(b) shows that all the couplings also increase in magnitude as the fine-structure constant $\kappa_0$ increases, and $g_{\perp z}$ is the most dominant one. A strongly attractive $g_{\perp z}$ favors Kekul\'e distorted (KD) state and lead to the appearance of KD states at small field (large $l_B$) or strong $\kappa_0$ as shown in our main results Fig.~1 and 2.
 

\section{Self-consistent Hartree-Fock Theory}\label{sec:HF section}

We now apply the previously determined couplings from the RG equations (Eqs.~\eqref{eq:e-e contribution to gux}, \eqref{eq:e-e contribution to guz}, \eqref{eq:e-p contribution to gxz}, \eqref{eq:e-p contribution to gzx}) to determine the ground state using self-consistent Hartree-Fock calculations.  

The single-particle basis we consider is given by the following:

\begin{align}\label{eq:LL wavefunction}
|nk\xi v s\rangle &= \frac{1}{\sqrt{2}^{1-\delta_{n0}}} \begin{pmatrix}
    \phi_{n,k} \\
    (1 - \delta_{n0})\xi \phi_{n-1,k}
\end{pmatrix}_{\Bar{A}\Bar{B}}\otimes |vs\rangle \\
\phi_{n,k}(\vec{r})&=\frac{1}{2^{n/2}\sqrt{n!l_B}\pi^{1/4}}H_{n}(\frac{x-kl_B^2}{l_B})e^{-\frac{(x-kl_B^2)^2}{2l_B^2}}\frac{e^{i k y}}{\sqrt{L_y}}
\end{align}
where $n$ the Landau level index take positive integer values, $0,1,2,3...$, $\xi=\pm1$ denotes the particle ($\xi=+1$) or hole ($\xi=-1$) state, $k$ is the momentum, $v$ is the valley and $s$ is the spin. $H_n$ is Hermite polynomials and $L_y$ is the size of the sample in the y-direction.  We will restrict our analysis to translation-invariant solutions, i.e.~the density matrix is independent of $k$, 
\begin{align}\label{eq:density matrix elements}
\langle nk\xi v s | \rho | n'k'\xi' v' s' \rangle \equiv  \rho_{n\xi,n'\xi^\prime}^{vv^\prime,ss^\prime}\,\delta_{k,k^\prime}.
\end{align}
With this assumption, we arrive at the following simplified Hartree-Fock Hamiltonian,
\begin{equation} \label{eq:H_{HF}}
        H_{HF} = T +  \Sigma^{C} + \Sigma^{\diamond} + \Delta_Z + \Delta_{\text{sub-latt}}
\end{equation}
Here $T $ represents the kinetic energy term for relativistic electrons in Landau levels, with a renormalized Fermi velocity computed at the magnetic $l_B$:
\begin{equation}
    T_{n\xi vs,n^\prime\xi^\prime v^\prime s^\prime} = \frac{\sqrt{2n}\hbar v_F}{l_B} \xi  \; \delta_{nn^\prime} \delta_{\xi\xi^\prime} \delta_{vv^\prime} \delta_{ss^\prime}.
\end{equation}
We note that $T$ is spin and valley independent and particle-hole symmetric. $\Sigma^{C}$ represents the self-energy derived from the long-range Coulomb potential $V_C=\frac{e^2}{4\pi\epsilon_0\epsilon_r|r-r'|}$.  We absorb the Hartree component of the self-energy into the zero energy, leaving the Fock component:

\begin{widetext}
\begin{align} \label{eq:Coulomb self energy}
    \Sigma^{C}_{n\xi vs, n^\prime \xi^\prime v^\prime s^\prime} &= 
    -\sum_{\substack{n_2 k\xi_2v_2s_2 \\ n_3 k^\prime\xi_3v_3s_3}} \langle nk\xi v s, n_2 k^\prime \xi_2 v_2 s_2 | V_c | n_3 k^\prime \xi_3 v_3 s_3, n^\prime k \xi^\prime v^\prime s^\prime \rangle \rho_{n_3\xi_3,n_2\xi_2}^{v_3v_2,s_3s_2}\\
    & =  -\sum_{n_2\xi_2n_3 \xi_3} V_{n\xi,n_2\xi_2,n_3\xi_3,n^\prime \xi^\prime}
    \rho_{n_3\xi_3,n_2\xi_2}^{vs,v's'}\\
    V_{n_1\xi_1,n_2\xi_2,n_3\xi_3,n_4 \xi_4}& = \sum_{k,k^\prime}\langle n_1k\xi_1, n_2 k^\prime \xi_2| V_c | n_3 k^\prime \xi_3, n_4 k \xi_4 \rangle 
\end{align}
In the second line, we simplify the expression by using the fact that the Coulomb potential is independent of spin and valley, and we perform integration over the momentum indices $k,k'$. Next, we address the short-range potential between two particles, represented as $V^s_{\alpha\beta} = g_{\alpha\beta}(\tau^\alpha \sigma^\beta)_1\delta(r-r') (\tau^\alpha \sigma^\beta)_2$ with $g_{\alpha\beta}$ being the renormalized short-range coupling constants. This short-range potential generates both the Hartree and Fock self energy:
\begin{align}
    \Sigma_{n\xi vs,n^\prime \xi^\prime v^\prime s^\prime}^{\diamond} &= \sum_{\alpha,\beta} \sum_{\substack{n_2 k\xi_2v_2s_2 \\ n_3k^\prime\xi_3v_3s_3}} \Big( \langle n k \xi v s, n_2 k^\prime \xi_2 v_2 s_2 | V^s_{\alpha \beta} | n^\prime \xi^\prime k v^\prime s^\prime, n_3 k^\prime \xi_3 v 3 s_3 \rangle \notag \\
    &\quad - \langle n k \xi v s, n_2 k^\prime \xi_2 v_2 s_2 | V^s_{\alpha \beta} | n_3 k^\prime \xi_3 v_3 s_3, n^\prime \xi^\prime k v^\prime s^\prime \rangle \Big) \rho_{n_3\xi_3,n_2\xi_2}^{v_3v_2,s_3s_2}.
\end{align}
\end{widetext}

In addition to these interaction terms, we also account for single-particle terms such as the Zeeman energy $\Delta_Z$ and potential energy difference between the two carbon sublattices generated by an aligned substrate $ \Delta_{\text{sub-latt}}$:
\begin{align}
     \Delta_Z &= -\epsilon_Z I_{KK^\prime} \otimes I_{\Bar{A}\Bar{B}} \otimes s^{z}_{ss^\prime}\\
    \Delta_{\text{sub-latt}}&=-\Delta_{BN} \tau^{z}_{KK'} \otimes \sigma^{z}_{\Bar{A}\Bar{B}} \otimes I_{ss'}
\end{align} 
where $\epsilon_Z=\mu_B |B|$ and we assume the 
g-factor to be $2$. $\Delta_{BN}$ is the strength of the sublattice-potential.
 We note in our basis $(\psi_{KA},\psi_{KB},\psi_{K'B},-\psi_{K'A})$, the $\Delta_{\text{sub-latt}}$ operator is antisymmetric in valley space.

Eq.~\eqref{eq:H_{HF}} is solved self-consistently using standard algorithms. To explore the possible ground states, we initialize the calculations with a variety of seeds, including random configurations, to construct the phase diagram displayed in Fig.~1 and Fig.~2. We did not find any coexistent phases \cite{das2022coexistence} despite keeping a significant of Landau levels, likely due to the delta-function approximation of the short-range interactions.

\subsection{Density Matrix of isotropic Dirac Fluid in a Magnetic Field}

The solution of the self-consistent Hartree-Fock equation show that the converged density matrix is diagonal with respect to the Landau level index:
\begin{equation} \label{eq:rho_diagonal}
\rho_{n\xi,n^\prime \xi^\prime}^{vv^\prime,ss^\prime} = \rho_{n\xi,n\xi^\prime}^{vv^\prime,ss^\prime} \, \delta_{n,n^\prime}. 
\end{equation}
This property arises because the converged solution describes an isotropic relativistic fluid under a magnetic field. As pointed out in Ref.~\cite{doi:10.1142/S0217979209062104}, the relativistic Landau-level form factors permit particles and holes with the same Landau level index, i.e., $\ket{n, \xi=+1}$ and $\ket{n, \xi=-1}$, to mix in an isotropic Dirac fluid. However, states with differing Landau level indices cannot mix, as their wavefunctions, characterized by distinct Hermite polynomials, lead to a spatially anisotropic density $n(\vec{r})\neq n(|\vec{r}|)$. Since our delta-function short-range potential is also isotropic, this property continues to hold, although we suspect that as interaction becomes too strong, the isotropic Dirac fluid may transition to a nematic phase, whose ground state possesses a finite electric multipole moment. In such scenario, Eq.~\eqref{eq:rho_diagonal} may no longer be valid.



This property enables the decomposition of the density matrix for an isotropic Dirac fluid into two components: the zeroth Landau level, denoted by $\hat{\rho}_{0}$ and the Dirac sea component $\rho_{DS}$,
\begin{equation}
    \rho = \hat{\rho}_{0} \oplus \rho_{DS},
\end{equation}
where the Dirac sea density matrix can be further represented as an infinite direct sum of higher Landau levels:  
\begin{equation}
    \rho_{DS} = \bigoplus_{n=1}^{\infty} \check{\rho}_n.
\end{equation}
Here $\hat{\rho}_{0} $ is a 4 by 4 matrix in spin-valley space while $\check{\rho}_n$ for $n\neq0$ is an 8 by 8 matrix, reflecting the mixing of particle and hole states:

\begin{align} \label{eq:nn_decouple}
    \check{\rho}_n &= \begin{pmatrix}
    \hat{\rho}_{n(-1),n(-1)} & \hat{\rho}_{n(-1),n(+1)}\\
        \hat{\rho}_{n(+1), n(-1)} & \hat{\rho}_{n(+1),n(+1)}.
    \end{pmatrix}
\end{align}
Here each of the block is a 4 by 4 matrix in spin-valley space. They are related to each others as follows: $\hat{\rho}_{n(-1),n(+1)}=\hat{\rho}_{n(+1),n(-1)}^\dagger$ and $\hat{\rho}_{n(+1),n(+1)}+\rho_{n(-1),n(-1)}=\mathbb{I}$, where $\mathbb{I}$ is a 4 by 4 identity matrix in spin-valley space.

In the absence of single-particle splittings $\Delta_Z=\Delta_{BN}=0$, we identified four solutions that align
with the phases in Kharitonov's original phase diagram -- Ferromagnetic (F), Antiferromagnetic (AF), Charge Density Wave (CDW), and 
Kekul\'e Distortion (KD), and their order paramters are shown in 
Table~\eqref{table:order parameter}.  For these states, each component of the density matrices is parameterized by just two elements: one proportional to the identity in the spin-valley space,  $\mathbb{I}$, and the other proportional to the order parameter $\mathcal{O}$:


\begin{align}
    \hat{\rho}_{0}&= \frac{1}{2}\left( \mathbb{I} + \mathcal{O} \right) \\
    \hat{\rho}_{n(-1),n(-1)}&= \alpha_n  \mathbb{I} + \beta_n \mathcal{O} \\
    \hat{\rho}_{n(+1),n(-1)}
    &= a_n \mathbb{I} + b_n \mathcal{O} 
\end{align}
where $n=1,2,3..$ is the set of positive integers (exclude zero).
We note that $\hat{\rho}_0$ retains the same form as when only the zeroth Landau level is considered, due to the property of Eq.~\eqref{eq:rho_diagonal}.
Note that the coefficients that appear in the above expressions are not independent because it is a single Slater-determinant $\rho^2=\rho$. Due to the block-diagonal structure, $\check{\rho}_{n}^2 = \check{\rho}_n$, thus

\begin{align}
\alpha_n^2+\beta_n^2+a_n^2+b_n^2 = \alpha_n\\
     2\alpha_n \beta_n+2a_nb_n = \beta_n.
\end{align}
Given these constraints, the coefficients can be parameterized using two independent variables, $\theta_n$ and $\phi_n$, which fully capture the degrees of freedom in the $n$-th component.
\begin{align}
    &\alpha_n = \frac{1}{2}+\frac{1}{4}(\cos{\theta_n}+\cos{\phi_n}) \,\ , \,\
    \beta_n = \frac{1}{4}(\cos{\theta_n}-\cos{\phi_n})\\
    & a_n = \frac{1}{4}(\sin{\theta_n}+\sin{\phi_n})\,\ , \,\ b_n = \frac{1}{4}(\sin{\theta_n}-\sin{\phi_n})
\end{align}
Our numerical results indicate that $\phi_n$ is approximately equal to $-\theta_n$. Consequently, we set $\phi_n = -\theta_n$, leading to a converged density matrix parameterized by a single angle:
\begin{align}\label{eq:density parametrization}
    \hat{\rho}_0 &= \frac{1}{2}(\mathbb{I}+\mathcal{O}) \\
    \hat{\rho}_{n(-1),n(-1)} &\approx \cos^2{\frac{\theta_n}{2}}\mathbb{I} \\
    \hat{\rho}_{n(+1),n(+1)} &\approx \sin^2{\frac{\theta_n}{2}}\mathbb{I} \\
    \hat{\rho}_{n(-1),n(+1)} & \approx \frac{1}{2}\sin{\theta_n} \mathcal{O}
\end{align}

\begin{figure}
    \centering
    \includegraphics[width=1\linewidth]{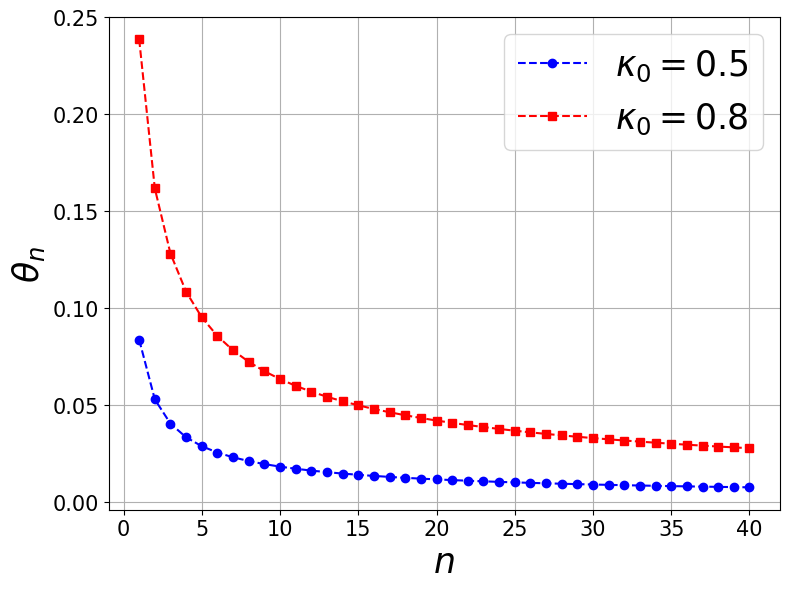}
    \caption{$\theta_n$ v.s. Landau level index $n$  for $\kappa_0=0.5$ and  $\kappa_0=0.8$. The angle $\theta_n$ is a single parameter that characterizes the effect of Landau level mixing, see Eq.\ref{eq:density parametrization}.}
    \label{fig:thetan vs n}
\end{figure}

These equations carry the following interpretation. The spin-valley order parameter $\mathcal{O}$ of the zeroth Landau level induces self-energy in remote Landau levels $\Sigma_{n\neq0}$. This self-energy leads to polarization of electrons in the remote Landau levels \cite{lukose2016coulomb}, aligning them along the spin-valley direction of the zeroth Landau level, $\hat{\rho}_{n(-1),n(+1)} \approx \frac{1}{2}\sin{\theta_n} \mathcal{O}$. Due to the isotropic condition, the eight-component wavefunctions within any given Landau level must be orthonormal to each other. Consequently, this polarization effect also induces a slight mixing of Dirac sea electrons ($\xi=-1$) with the unoccupied Landau levels ($\xi=1$), $\hat{\rho}_{n(-1),n(-1)} = \cos^2{\frac{\theta_n}{2}}\mathbb{I}$ and $
    \hat{\rho}_{n(+1),n(+1)} =\sin^2{\frac{\theta_n}{2}}\mathbb{I}$. We note that $\sin(\theta_n)$ is related to orbital magnetic moments of the isotropic Dirac fluid in a magnetic field, reflecting the magnitude of the Haldane gap which has opposite signs in opposite valleys, i.e.~ $N_{KA}-N_{KB}+N_{K'B}-N_{K'A}=\frac{1}{2}\sum_n \sin(\theta_n)$. Fig.~\eqref{fig:thetan vs n} shows that $\theta_n$ decays slowly with $n$. 



\begin{table}[t]
\centering
\begin{tabular}{|c|c|c|}
\hline
 Symmetry-broken State &  $\mathcal{O}$ & OP symmetry\\ \hline
Antiferromaget (AF)  & $\tau^z s^z$  &  $ U(1)_{KK'}\otimes SU(2)_{ss^\prime}$ \\ \hline
Kekul\'e Distorted (KD)   & $\tau^x s^0$  & $U(1)_{KK^\prime} \otimes SU(2)_{ss^\prime}$\\ \hline
Charge-density wave (CDW)   & $\tau^z s^0$ & $Z_{2KK^\prime} \otimes SU(2)_{ss^\prime}$    \\ \hline
Ferromagnet (F)   & $\tau^0 s^z$ &$U(1)_{KK^\prime} \otimes SU(2)_{ss^\prime}$ \\ \hline
Canted Antiferromagnet (CAF)   &  $\tau^0 s^z$,$\tau^z s^x$   & $U(1)_{KK^\prime} \otimes U(1)_{ss^\prime}$ \\ \hline
Canted Kekul\'e Distorted (CKD)   & $\tau^z s^0$,$\tau^x s^0$   & $U(1)_{KK^\prime} \otimes SU(2)_{ss^\prime}$ \\ \hline
\end{tabular}
\caption{Symmetry broken states with their corresponding order parameters and symmetries.}
\label{table:order parameter}
\end{table}

\subsection{Magnetic Anisotropic Energy from the Dirac Sea and Zeroth Landau Level}

In this subsection, we discuss the magnetic anisotropic energy contributions from zero LL and Dirac sea. First, we note that the self-energy, due to the isotropic condition defined in Eq.~\eqref{eq:rho_diagonal}, must also be diagonal with respect to the Landau level index. This allows us to express the both the spin-valley-isotropic self-energy ($\Sigma^C$) and spin-valley-anisotropic self-energy ($\Sigma^{\diamond}$) as the following form,
\begin{align} \label{eq:Sigma_block_separation}
    \Sigma^C = \hat{\Sigma}_{0}^C \oplus \Sigma_{DS}^C \;\;,\;\;
    \Sigma^{\diamond} = \hat{\Sigma}_{0}^{\diamond}\oplus \Sigma_{DS}^{\diamond},
\end{align}
where the Dirac sea contribution can be further represented as an infinite direct sum from higher Landau levels:  
\begin{equation}
    \Sigma_{DS}^C = \bigoplus_{n=1}^{\infty} \check{\Sigma}_n^C \;\;,\;\;
    \Sigma_{DS}^{\diamond} = \bigoplus_{n=1}^{\infty} \check{\Sigma}_n^{\diamond}.
\end{equation}
Importantly, because of the delta-function potential, the spin-valley anisotropic self-energy $\check{\Sigma}_n^{\diamond}$ does not decay with Landau level $n$: 
\begin{equation}
\check{\Sigma}_n^{\diamond}=\check{\Sigma}^{\diamond}
\end{equation}
The explicit formula for the self-energy is rather long and we provide them in the appendix. 

\begin{figure}[t]
    \centering
    \includegraphics[width=1\linewidth]{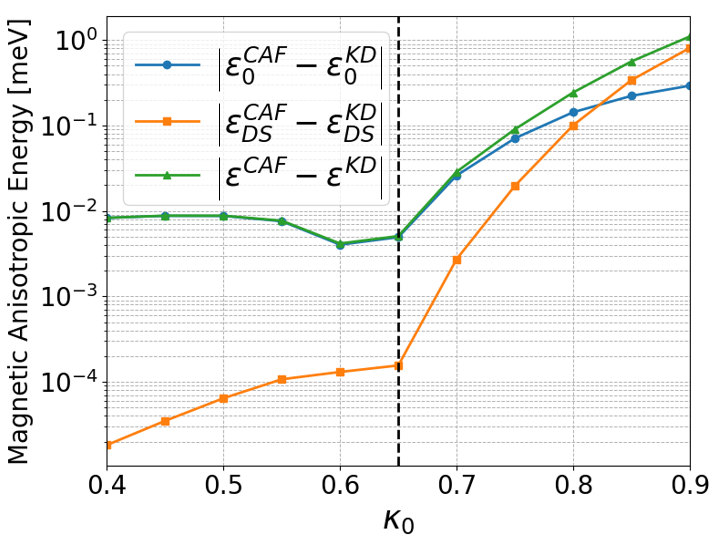}
    \caption{Absolute energy difference between the canted antiferromagnet (CAF) and Kekule-distorted (KD) states as a function of $\kappa_0$, including contributions from the Dirac sea and the zeroth-Landau level. The black dashed line indicates the critical $\kappa_0$ where an energy level crossing occurs. The absolute energy difference is plotted on a logarithmic scale to highlight the relative dominance of the zeroth-Landau level contribution at small $\kappa_0$ and the Dirac sea contribution at larger $\kappa_0$. Here $B=10$ T.} 
    \label{fig:magnetic anisotropic energy}
\end{figure}

Using the block-diagonal structure of Eq.~\eqref{eq:Sigma_block_separation}, the Hartree-Fock Hamiltonian can be decomposed into contributions from the zeroth Landau level and the Dirac sea:  $H_{HF}=H_{HF,0} \oplus H_{HF,DS}$. This Dirac-sum decomposition allows us to separate the zeroth Landau-level and Dirac sea contributions to the total energy per particle of a converged density matrix $\rho$:
\begin{align}
    \epsilon&=\frac{E}{N}=\frac{1}{2N}\text{Tr}[(H_{HF}+T)\rho] \nonumber \\
    &= \frac{\text{Tr}[(H_{HF,0}+T_0)\rho_0]}{2N} + \frac{\text{Tr}[(H_{HF,DS}+T_{DS})\rho_{DS}]}{2N} \nonumber \\
    &\equiv \epsilon_{0} +  \epsilon_{DS}
\end{align}
Here $N=\text{Tr}(\rho)$. The energy difference between two symmetry-broken ordered states, can be understood as spin valley expanded magnetic anisotropic energy. Similar to conventional itinerant electron ferromagnets, there is significant exchange energy associated with collective spin alignment, but the energy required to align the spin-polarization to a specific direction, i.e., the magnetic anisotropic energy, is comparatively small.  Fig.~\ref{fig:magnetic anisotropic energy} shows the magnetic anisotropic energy difference between CAF and KD states as function of $\kappa_0$. Here $\epsilon_0^{KD}$ and  $\epsilon_{DS}^{KD}$ represent the zeroth Landau-level and Dirac sea contributions to the total energy of the KD state. Similarly, $\epsilon_0^{CAF}$ and  $\epsilon_{DS}^{CAF}$ 
are the zeroth Landau-level and Dirac sea contributions to the total energy of the CAF state.
We found that at small $\kappa_0$, the magnetic anisotropic energy primarily arises from the zeroth Landau level, with negligible contribution from the Dirac sea, resulting in CAF as the groundstate.
However, as $\kappa_0$ increases, a phase transition from the CAF state to the KD state occurs between \(\kappa_0 = 0.6\) and \(\kappa_0 = 0.65\), as indicated by the vertical black dashed line. When \(\kappa_0 > 0.8\), the Dirac sea contribution surpasses that of the zero Landau level. When $\kappa_0$ is large, the particle-hole components of the remote density matrix is also large (see Fig.~\ref{fig:thetan vs n}) and this leads to a more pronounced self-energy difference between the CAF and KD states. Our results show that for large $\kappa_0$, contributions from the Dirac sea must be treated in a non-perturbative framework to accurately capture the magnetic-anisotropic energy landscape.

\subsection{Zeorth Landau Level Wavefunctions}

\begin{figure}[t]
    \centering
    \includegraphics[width=1\linewidth]{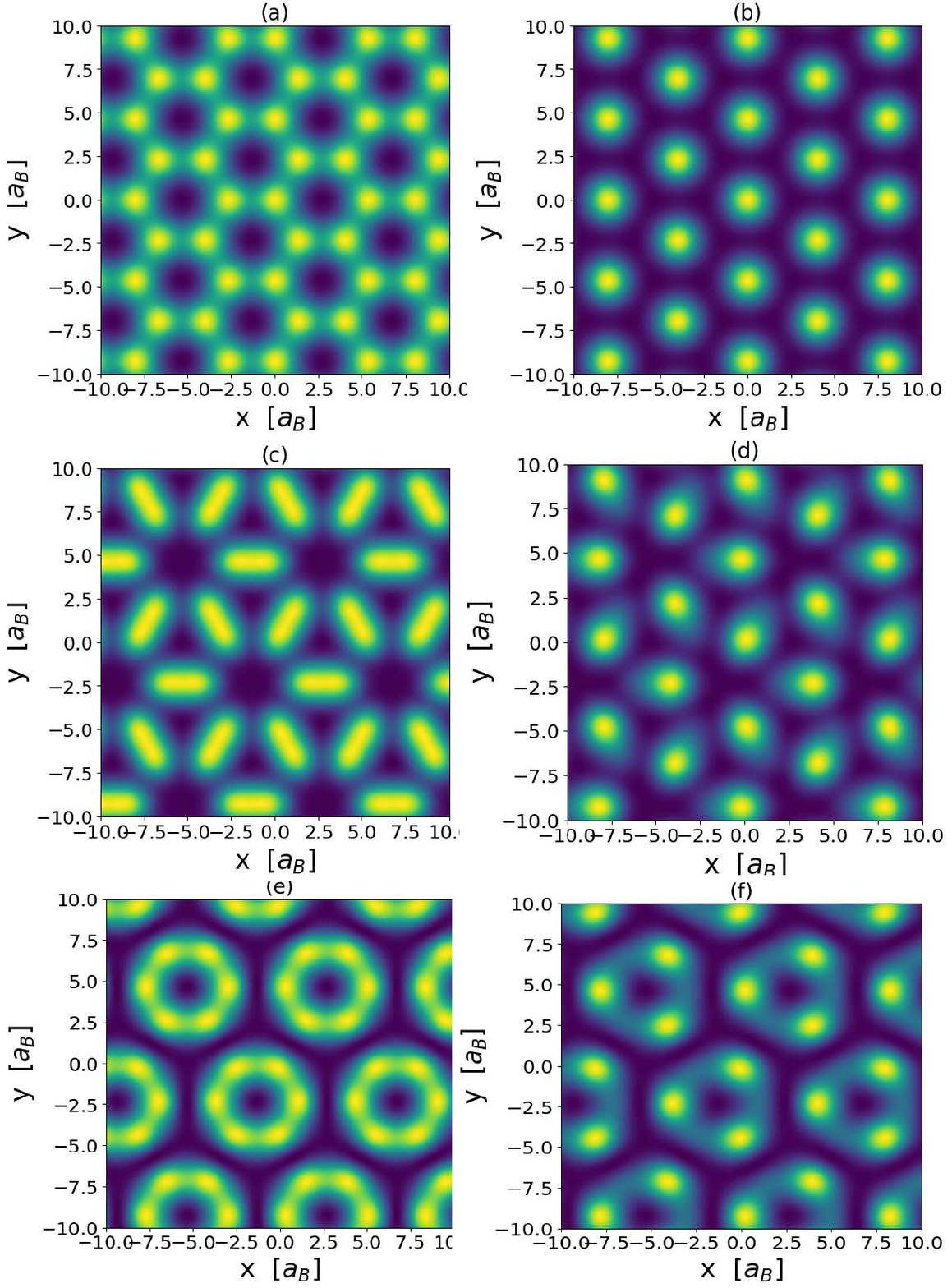}
    \caption{Electron density distribution $|\Psi|^2$ for the zeroth Landau level wavefunctions at various locations in Fig.~\ref{fig:zero_LL_phase_diagram}. Here the bright colors indicate regions of peak density and length scale is in Bohr radius $a_B$. 
    a) AF state at \(\kappa_0 = 0.3\),  \(B = 10\, \text{T}\), $\Delta_{BN}=0$
    b) Sublattice-polarized state at \(\kappa_0 = 0.3\) , \(B = 5\, \text{T}\), $\Delta_{BN}=1\,\text{meV}$.
    c) KD state when the relative phase angle between the two valleys $\phi=0$ at \(\kappa_0 = 0.7\), \(B = 10\, \text{T}\) and  $\Delta_{BN}=1\,  \text{meV}$. d) Canted-KD state at \(\kappa_0 = 0.2\) and \(B = 5\, \text{T}\) and $\Delta_{BN}=1\, \text{meV}$ 
    e) KD state with $\phi=\pi$ at $\kappa_0=0.7$,$B=10\, \text{T}$ and  f) Canted KD state at \(\kappa_0 = 0.2\) and \(B = 5\, \text{T}\) with $\Delta_{BN}=1\, \text{meV}$.
}
    \label{fig:electron-wavefuntion}
\end{figure}

\label{sec:wavefunction}
In this section, we present the zeroth Landau level electron wavefunctions at various points within the phase diagram shown in Fig.~\eqref{fig:zero_LL_phase_diagram}. These wavefunctions provide a direct means of comparison with recent STM experimental studies \cite{Coissard2021ImagingTQ,doi:10.1126/science.abm3770,PhysRevB.100.085437}. The main finding of our analysis is that when the relativistic quantum fluid in a magnetic field is isotropic, it imposes a stringent limitation on the wavefunction: the zeroth Landau level wavefunction cannot mix with higher Landau levels. Under this isotropic condition, the range of many-body states observable in STM experiments is severely restricted, with some representative pattern shown in Fig.~\ref{fig:electron-wavefuntion}. Therefore, the detection of more intricate wavefunctions near charge neutrality in STM measurements would likely indicate mixing between the zeroth Landau level and higher Landau levels. Such wavefunction mixing could be induced, for instance, by an external moir\'e potential arising from the substrate.


From the $K\cdot p$ approximation, the three-dimensional electron wavefunction, $\Psi^{(0)}(x,y,z)$, can be represented as a product of two wavefunctions: the slowly varying part ($\psi$) derived from the SCHF calculations, and a rapidly varying part ($u$):
\begin{align}
&\Psi^{(0)}(x,y,z) = \sum_{n,\tau,\sigma,s} \psi^{(0)}_{n\tau \sigma s}(x,y) u_{\tau \sigma}(x,y,z).
\end{align}
Here, $\psi_{n\tau\sigma s}^{(0)}$ is the wavefunction localized in sublattice $\sigma$, and it can be written as linear combinations of the particle and hole components of the SCHF eigenfunctions:
\begin{align}
&\psi^{(0)}_{n K A s}(x,y) \notag \\
&= \frac{  \langle x,y | n(+) K s \rangle c^{(0)}_{n(+) K s} + \langle x,y | n(-) K s \rangle c^{(0)}_{n(-) K s}   }{\sqrt{2}^{1+\delta_{n0}}} \label{eq:eigen wavefunction in SCHF basis}\\
&\psi^{(0)}_{n K' B s}(x,y) \notag \\
&= \frac{ \langle x,y | n(+) K' s \rangle c^{(0)}_{n(+) K' s} + \langle x,y | n(-) K' s \rangle c^{(0)}_{n(-) K' s} }{\sqrt{2}^{1+\delta_{n0}}}.  \label{eq:eigen wavefunction in SCHF basis2}
\end{align}
In the above expression, we have used sublattice-valley locking property of the zero Landau level to simplify the equations.
The complex-value coefficients $c^{0}_{n\xi\tau s}$ are the entries of the eigenvector of the mean-field Hamiltonians.
Importantly, since the zero Landau level does not mix with other Landau levels in an isotropic Dirac fluid, 
\begin{equation} \label{eq:0LL_weights}
    c^{(0)}_{n(\pm) \tau s}=0 \;\text{for}\; n\neq0.
\end{equation}
and $c^{(0)}_{0(-) \tau s}=c^{(0)}_{0(+) \tau s}$.  The 3D wavefunction \(u_{\tau \sigma}(x,y,z)\) can be approximated as a linear combination of 2$p_z$ atomic orbital wavefunctions \(\phi_{2p_z}(\vec{r})\):
\begin{align}
u_{\tau \sigma}(x,y,z) &= \sum_{\vec{R}} e^{i \tau \vec{K} \cdot \vec{R}_\sigma} \phi_{2p_z}(\vec{r} - \vec{R})\\
\phi_{2p_z}(\vec{r}) &= \frac{2Z^2\sqrt{Z}}{\sqrt{96a_B^5}} r e^{-\frac{Zr}{2a_B}} Y_0^1(\theta,\phi) 
\end{align}
Here, $Z$ is the effective nuclear charge and $\vec{R}$ represents all lattice vector. 

Since the wavefunction \( u_{\tau\sigma} \) changes quickly at the scale of lattice spacing, whereas \(\psi_{\tau\sigma s}\) varies over a much longer length scale set by the magnetic length \( l_B \), we can approximate \(\psi_{\tau\sigma s}\) as constant when considering changes over a few carbon lattice spacing. Under this approximation, the overall zeroth Landau level wavefunction is given by the following:

\begin{align}
   & \Psi^{(0)}(\vec{r}) \approx \sum_{s} \bigg[ \frac{1}{2} \left( c^{(0)}_{0(+)K s} + c^{(0)}_{0(-)K s} \right)u_{KA}(\vec{r})  \notag \\
    &+\frac{1}{2}\left( c^{(0)}_{0(+)K' s} + c^{(0)}_{0(-)K' s} \right)u_{K'B}(\vec{r}) \bigg]   \\
    & = \sum_s \left(c^{(0)}_{0(+)Ks}u_{KA}(\vec{r})+c^{(0)}_{0(+)K' s}u_{K'B}(\vec{r})  \right) 
\end{align}
where $\vec{r}=(x,y,z)$. Next, we will plot probablity density of the electron wavefunction $|\Psi^{(0)}(x,y,z)|^2$ for different ground states in the $B--\kappa_0$ phase diagram. We fix the \(z\)-direction fixed at \(2a_B\).
We first describe ground states that are easily identifiable in experiments, such as the CAF and the sublattice-polarized phases. We then discuss the more intriguing Kekul\'e bond phases. It is important to note that when the sublattice-polarized state is induced by an aligned Boron-Nitride (BN) potential rather than interaction effects, it may be misleading to label such a state as a charge-density wave. Therefore, we refer to it as the sublattice-polarized phase.

Fig.~\ref{fig:electron-wavefuntion}
(a) shows the electron density distribution in the canted antiferromagnetic (CAF) state. In this state, electrons have an equal probability of occupying both sublattices (valleys), yet their spins are not collinear. The spin quantization axes are nearly aligned in the xy-plane and oriented oppositely, but due to finite Zeeman energy, they exhibit a slight tilt out of the plane. Bright areas highlight regions of peak charge density, which has a honeycomb pattern, respecting the underlying lattice symmetry of graphene. The CAF state is adiabatically connected to the ferromagnetic (FM) state, where all spins align in the direction dictated by the Zeeman field.

Fig.~\ref{fig:electron-wavefuntion}
b) shows the sublattice-polarized phase, where electrons in the zeroth Landau level are confined to the same sublattice but have opposite spin projections. It has a $C_3$, as opposed to the usual $C_6$ symmetry observed in FM and CAF.

Fig.~\ref{fig:electron-wavefuntion}(c-f)show a class of bond-ordered states that is also termed the Kekule-distorted state. This state is distinct from the above groudnstates as it breaks the original discrete translational symmetry of the graphene lattice. This is due to the SCHF eigenfunctions mixing two Bloch states from opposite valleys, which are distinct eigenvectors of the graphene's lattice translation operator, resulting in a threefold enlargement of the unit cell. The electron density distribution of these states vary depending on the relative phase $\phi$ (the azimuthal angle on the valley Bloch sphere, with a polar angle $\theta = \pi/2$) between the two valleys. 

Fig.~\ref{fig:electron-wavefuntion}c) shows  the groundstate for $\phi = 0$ where the resulting electron density peaks at the AB sublattice bond. This resembles a chemical bond maintains the $C_{6}$ symmetry of the graphene lattice. This is sometimes termed the Kekul\'e -O distorted state.
Fig.~\ref{fig:electron-wavefuntion}d)
shows the so-called canted KD state in the presence of the BN potential. The order parameter of this state tilts away from the north/south pole forming an angle of $\theta = \arccos{\langle \tau^{z}s^0 \rangle/2}$.
For $\kappa_0=0.2$, $B=10\, \text{T}$ and $\Delta_{BN}=1\, \text{meV}$, $\theta \approx \frac{7\pi}{26}$.
Fig.~\ref{fig:electron-wavefuntion} e-f) show another type of the Kekul\'e state and canted-Kekul\'e state when the relative phase angle $\phi=\pi$, which breaks the the sixfold rotational symmetry $C_{6}$  of the graphene lattice. The polar angle $\theta$ of these two states are the same due to the  $U(1)_{KK'}$ symmetry of the valley.

\begin{figure}
    \centering
    \includegraphics[width=1\linewidth]{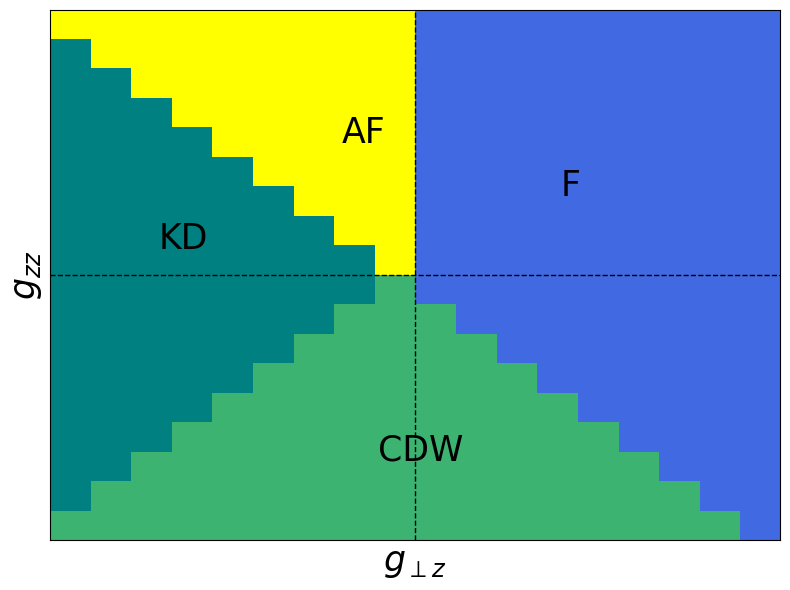}
    \caption{The original $(u_\perp-u_z)$ Kharitonov phase diagram \cite{kharitonov2012phase} obtained via zeroth-Landau level projection. Here $u_{z} = g_{zz}/(2\pi l_B^2)$ and $u_{\perp} = g_{\perp z}/(2\pi l_B^2)$, while terms proportional to $g_{\perp \perp}$ and $g_{z\perp}$ vanish due to zeroth Landau level projection.}
    \label{fig:kharitonov diagram}
\end{figure}

\begin{figure}[t]
    \centering
    \includegraphics[width=1.1\linewidth]{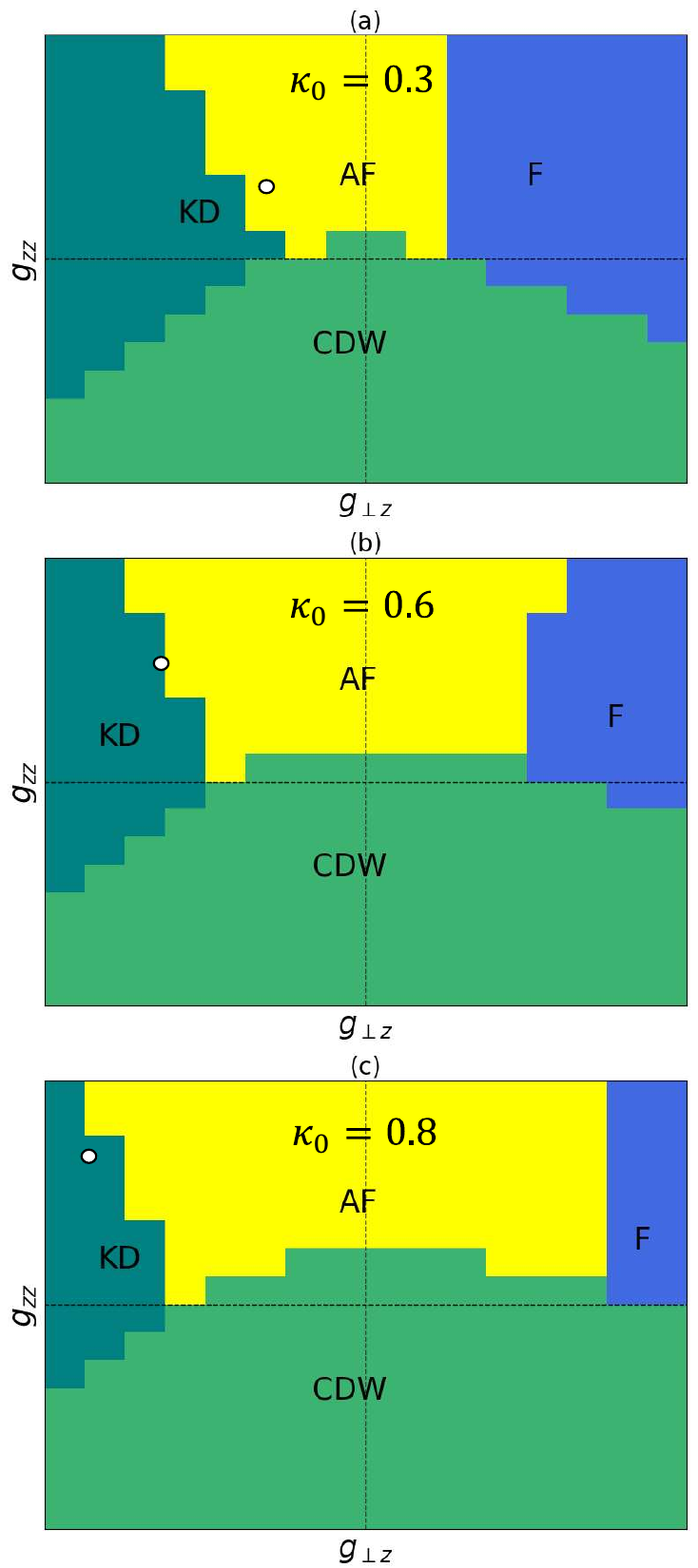}
    \caption{Evolution of the $\nu=0$ Kharitonov phase diagram at different values of $\kappa_0$ \textit{without} zeroth-Landau level projection. To visualize the Kharitonov phase diagram at different $\kappa_0$, we perform self-consistent Hartree-Fock calculations as a function of $g_{zz}$ and $g_{\perp z}$ for a given $(g_{z\perp}(\kappa_0), g_{\perp \perp}(\kappa_0), v_F(\kappa_0))$. The white dot represents the actual renormalized values, $g_{zz}(\kappa_0)$ and $g_{\perp z}(\kappa_0)$,  in the $g_{\perp z} - g_{zz}$ parameter space. Panel (a) corresponds to $\kappa_0 = 0.3$, panel (b) to $\kappa_0 = 0.6$, and panel (c) to $\kappa_0 = 0.8$. All diagrams are generated with a magnetic field of $B = 10$ T without Zeeman effect and sublattice potential from aligned substrate.}
    \label{fig:phase diagram in coupling space}
\end{figure}

\subsection{Kharitonov Phase Diagram}

In this section, we discuss why it is tricky to identify the $\nu=0$ ground state at various values of the fine-structure constant $\kappa_0 = e^2/\hbar v^0_F \epsilon_0\epsilon_r$ within the Kharitonov phase diagram.  When the short-range coupling constants--$g_{\perp\perp}$, $g_{\perp z}$, $g_{z\perp}$, and $g_{zz}$--are projected onto the zeroth Landau level, the sublattice-valley locking structure preserves only those coupling constants that flip both the valley and sublattice at the same time, Consequently, only $g_{zz}$ and $g_{\perp z}$ are non-zero, c.f. Fig.~\ref{fig:short-range coupling diagram}.  Following this projection, the $\nu=0$ phase diagram is constructed using the normalized coupling constants: $u_\perp \equiv g_{\perp z}/(2\pi l_B^2)$ and $u_z \equiv g_{zz}/(2\pi l_B^2)$. Both $u_z$ and $u_\perp$ are proportional to the magnetic field strength $B$, and their ratio to the Zeeman energy, independent of $B$, typically estimated from experimental data to be between 5 to 10 times.
Fig.~\eqref{fig:kharitonov diagram} shows the original  Kharitonov phase diagram without Zeeman and Boron Nitride potential. Note that the phase boundary between the AF and KD states lies along a 135-degree line in the second quadrant.

In Fig.~\eqref{fig:phase diagram in coupling space}a)-c) we present the phase diagram as a function of $g_{zz}$ and $g_{\perp,z}$ without zeroth Landau-level projection, varying across different values of the fine-structure constant, $\kappa_0$. We perform a self-consistent Hartree-Fock calculations as a function of $g_{zz}$ and $g_{\perp,z}$ while keeping the other coupling constants that do not survive zeroth Landau level projection fixed at their renormalized values $(g_{\perp\perp},g_{\perp z}, v_F) =
(g_{\perp\perp}(\kappa_0),g_{\perp z}(\kappa_0), v_F(\kappa_0))$.
The specific renormalized values for $g_{zz}$ and $g_{\perp z}$ at each $\kappa_0$ are indicated by white dots in Fig.~\eqref{fig:phase diagram in coupling space}. The energy window for this phase diagram is chosen to be approximately \( 1200 \, \text{meV} \cdot \text{nm}^2 \) , and the discretization is \( 100 \, \text{meV} \cdot \text{nm}^2 \).

The phase diagrams reveal shifts in phase boundaries due to the influence of remote Landau levels compared to those predicted by the zeroth Landau level projected phase diagram. As $\kappa_0$ increases, the AF phase expands in the second quadrant while the KD phase shrinks.
However, the renormalization group solution for the short-range coupling constant is such that $ 
-g_{\perp z} > g_{zz} $ at strong $\kappa_0$ (~see Fig.~\ref{fig:RG flow of coupling}). Thus, at certain critical values of $\kappa_0$, the system transitions from the AF phase to the KD phase. Consequently, under weak screening conditions, the system favors the KD state, while the AF state is preferred under strong screening.
As we see from the above analysis, asserting the $\nu=0$ groundstate from just a two-dimensional parameter space of $g_{zz}$ and $g_{\perp z}$ can be quite challenging since the coupling constants undergo very strong renormalization.

\section{Conclusion}\label{sec:conclusion}

In this work, we have demonstrated a Landau-level mixing induced phase transition \cite{wei2024landau} for both the $\nu=0$ and $\nu=\pm1$ quantum Hall states. These transitions are not driven by single-particle effects such as the Zeeman field or Boron Nitride potential, but rather originate from the subtle change of valley-sublattice-dependent interactions mediated by the Dirac sea as the system navigates the $\kappa_0-B$ (fine-structure constant--magnetic field) parameter space. The Dirac sea has two important effect: First, the effective interactions mediated by states near the linear dispersion cutoff are substantial due to the high density of states, which significantly enhances the magnitude of the valley-sublattice-dependent interaction coupling constants. Second, the first-order exchange effects with the Dirac sea result in pronounced Landau-level mixing. This mixing changes the wavefunctions with the same Landau level index -- not merely causing energy shifts -- and contributes significantly to the magnetic anisotropic energy when $\kappa_0$ is order of one number. We addressed both of these aspects using a combination of renormalization group and self-consistent Hartree-Fock theory. Our study shows that the $\nu=0$ quantum Hall state transitions from a (canted) antiferromagnetic state to a Kekule distorted state as $\kappa_0$ increases (screening decreases) and magnetic field decreases. This result agrees with our earlier findings using second-order perturbation theory to compute the spin-valley effective interaction to leading order in $\kappa_0$. However, as noted in Ref.~\cite{wei2024landau}, predictions of magnetic field-induced phase transitions using second-order perturbation theory are unreliable, and our two-step approach provides a more accurate theoretical estimate of the critical magnetic field. For $\nu=\pm1$, we predict that the spin-polarized charge density wave state predominates most of the $\kappa_0-B$ phase diagram, even without boron nitride potential, transitioning to a Kekule distorted phase as $\kappa_0$ increases and magnetic field decreases. This transition is primarily due to the more rapid increase of $g_{\perp z}$ compared to $g_{zz}$ under renormalization group flow.

We expect that these findings may extend to 
fractional fillings \cite{Sodemann2013BrokenSS, An2024MagneticAL}. Currently, our renormalization group approach tracks only four coupling constants. However, to more accurately reflect the complexity of the interactions, it would be advantageous to develop a comprehensive program that includes coupling constants describing finite-range spin-valley dependent interactions. These interactions can emerge from delta-function potentials under RG flow just like the Berk-Schrieffer diagram \cite{berk1966effect} generates long-range spin-triplet attractions from delta-function type repulsion. Recent studies, such as those by Ref.~\cite{kousa2024orbital}, have also emphasized the importance of the Dirac sea in determining ideal conditions to spectrally isolate the $N=1$ Landau level to explore $\nu=5/2$ physics in bilayer graphene. Applying our two-step approach to bilayer graphene could provide new insights into its phase diagram under strong magnetic fields. 

\textit{Acknowledgment:} We thank Ganpathy Murthy,  Inti Sodemann, Oskar Vafek and Nemin Wei for discussions. 
\appendix

\section{Calculation of vacuum polarization}

Here, we discuss the calculation details of the vacuum polarization in Eq.~\eqref{eq:vacuum polarization} using the Feynman parameter method:

\begin{align}
    &\Pi(\vec{p}, p_0) =-N\int \frac{d^3q}{(2\pi)^3}\text{Tr}\left(G_0(p_0+q_0,\vec{p}+\vec{q}) G_0(q_0,\vec{q}) \right)\\
    & = -N\int \frac{d^3q}{(2\pi)^3}\text{Tr}\big(  \frac{i(p_0+q_0)I+v_F\vec{\sigma}\cdot(\vec{p}+\vec{q})}{(p+q)^2}\\
    &\times \frac{iq_0I+v_F\vec{\sigma}\cdot\vec{q}}{q^2}\big)\\
    & = -4N \int \frac{d^3q}{(2\pi)^3} \frac{-(p_0+q_0)q_0+v_F^2(\vec{p}+\vec{q})^2}{q^2(p+q)^2}
\end{align}
In the above equations, \( q^2 = q_0^2 + v_F^2\vec{q}^2 \). Next, we will apply the Feynman parameter method:

\begin{align}
    \frac{1}{AB} = \int_{0}^{1} dx \frac{1}{(xA+(1-x)B)^2}
\end{align}
and further change the variable to \( l = q + xp \),

\begin{align}
   &\Pi(\vec{p}, p_0) \\
   &= -4N \int \frac{d^3q}{(2\pi)^3}\int_0^{1} dx \frac{-(p_0+q_0)q_0+v_F^2(\vec{p}+\vec{q})\cdot\vec{q}}{((1-x)q^2+x(p+q)^2)}\\
          & = 4N\int_0^{1} dx \int \frac{d^3l}{(2\pi)^3} \frac{l_0^2-v_F^2\vec{l}^2-x(1-x)(p_0^2-v_F^2\vec{p}^2)}{(l^2+p^2(x-x^2))^2}\\
    & = \frac{4N}{v_F^2}\int_0^{1} dx \int \frac{d^3l}{(2\pi)^3} \frac{-\frac{1}{3}l^2-x(1-x)(p_0^2-v_F^2\vec{p}^2)}{(l^2+p^2(x-x^2))^2}
\end{align}
In the last equation above, we rescaled the in-plane components of \( l \). Due to the isotropy of the integration, we can replace \( \langle l_0^2 \rangle = \langle l_1^2 \rangle = \langle l_2^2 \rangle \) with \( \frac{1}{3}\langle l^2 \rangle \).The resulting integral can be evaluated using:

\begin{align}
    \int\frac{d^d l}{(2\pi)^d}\frac{1}{(l^2+\Delta)^n}&=\frac{\Gamma(n-d/2)}{(4\pi)^{d/2}\Gamma(n)}\frac{1}{\Delta^{n-d/2}}\\
        \int\frac{d^d l}{(2\pi)^d}\frac{l^2}{(l^2+\Delta)^n}&=\frac{d}{2}\frac{\Gamma(n-d/2-1)}{(4\pi)^{d/2}\Gamma(n)}\frac{1}{\Delta^{n-d/2-1}}
\end{align}
thus, the vacuum polarization is given by\cite{Son2007QuantumCP}:

\begin{align}
    \Pi(\vec{p}, p_0) &= \frac{4N}{v_F^2}\frac{\sqrt{\pi}}{(4\pi)^{3/2}}\frac{p^2-(p_0^2-v_F^2\vec{p}^2)}{p^2}\int_{0}^{1}dx \sqrt{x-x^2} \\
    & = \frac{N}{8}\frac{\vec{p}^2}{\sqrt{p_0^2+v_F^2\vec{p}^2}}
\end{align}

\section{Landau level cutoff dependence}
Here, we studied energy convergence of HF calculation with respect to the cutoff of Landau levels.
\begin{figure}[h!]
    \centering
    \includegraphics[width=1.0\linewidth]{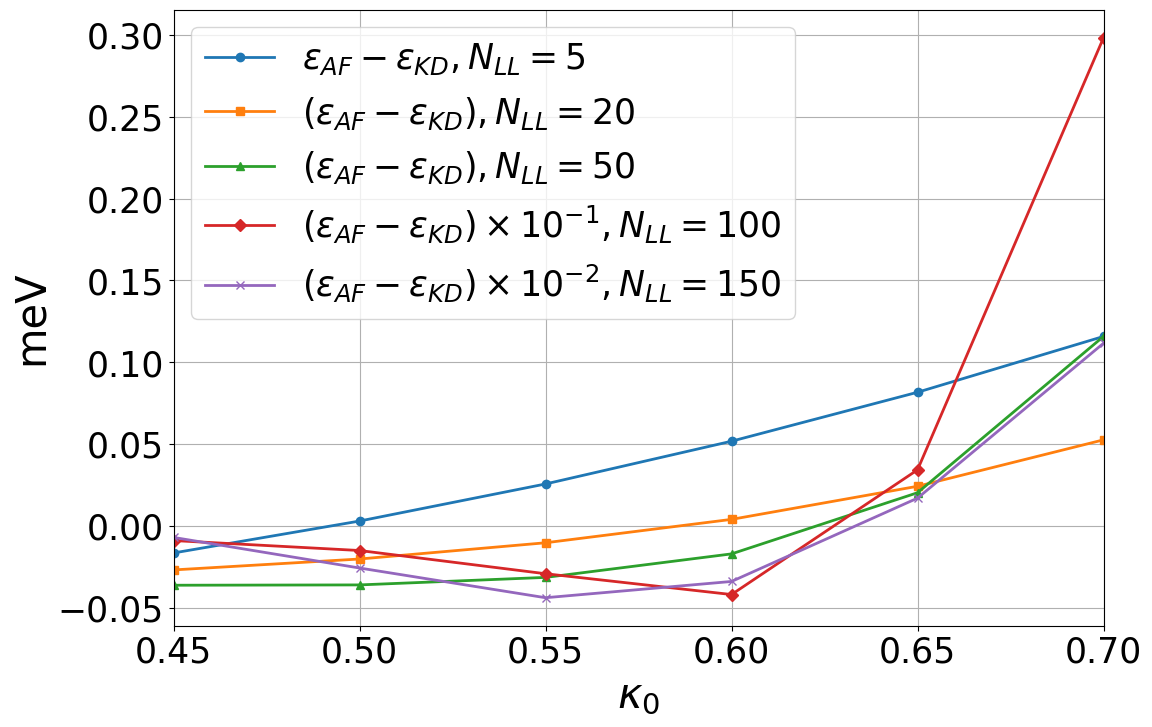}
    \caption{The energy difference (per particle) between AF state and KD state in the Hartree-Fock approximation at $B=10$T. Zeeman and sublattice polarization energies are set to zero.}
    \label{fig:energy_diff_at_diff_LL}
\end{figure}
Fig.~\eqref{fig:energy_diff_at_diff_LL} shows that at $N_{LL} = 5$, the phase transition between the KD and AF state occurs near \(\kappa_0 \approx 0.5\). As the cutoff increases to \(N_{LL} = 20\), the transition point shifts to around \(\kappa_0 \approx 0.6\). For larger cutoffs such as \(N_{LL} = 50\), \(100\), and \(150\), the first order phase transition point shows only a minor shift, while their energy difference becomes stronger. Thus, increasing $N_{LL}$ does not shift the phase transition point too much but makes the phase transition strongly first-order. For our Hartree-Fock calculations, we chose $N_{LL}=50$ as a balanced compromise between convergence and computational efficiency.

\section{Explicit formula for the Self Energies}
In this appendix, we discuss in detail the formulas for various self-energies, discussing their derivation and physical implications.
\begin{widetext}

\subsection{Kinetic energy}
With the parametrization of the density matrix in Eq.~\eqref{eq:density parametrization}, the total kinetic energy is given by:
\begin{align}
E_T =\text{Tr}(\hat{T} \rho) = -\sum_{n \geq 0}^{N_{LL}} \frac{\sqrt{2} \hbar \omega_c}{l_B^2} \sqrt{n} \cos{\theta_n}.
\end{align}
The factor $\cos{\theta_n}$ arises due to the LL mixing. This mixing effect increases the energy, as the wavefunction can extend into the positive Landau levels.The total kinetic energy diverges due to the large number of particles in the Dirac sea. To cancel the divergence due to large number of particles, one can calculate the average energy instead. Alternatively, if the focus is solely on the LL mixing effect, the divergence can be addressed by subtracting the contribution from the background density, which is introduced in the next section. This subtraction effectively cancels out the large divergence and isolates the energy associated with LL mixing.

\subsection{Self-energy of long-range Coulomb potential}

In Eq.~\eqref{eq:Coulomb self energy}, $\Sigma^{C}_{n\xi,n^\prime\xi^\prime}$ is contributed by the following matrix elements $V_{n\xi,n_2\xi_2,n_3\xi_3,n^\prime\xi^\prime}$. Their explicit form is given by the following:

\begin{align}
V_{n_1\xi_1,n_2\xi_2;n_3\xi_3,n_4\xi_4}
 &=\sum_{k,k^\prime}\langle n_1k\xi_1, n_2 k^\prime \xi_2| V_c | n_3 k^\prime \xi_3, n_4 k \xi_4 \rangle \\
&=\sum_{k,k^\prime}\int d^2r_1d^2r_2V_c(r_1-r_2)
\left[ \psi^{\dagger}_{n_1k\xi_1}(r_1)\cdot \psi_{n_3k^\prime\xi_3}(r_1) \right] \left[ \psi^{\dagger}_{n_2k^\prime\xi_2}(r_2)\cdot \psi_{n_4k\xi_4}(r_2) \right]\\
& =\sqrt{2}^{\sum_i\delta_{n_i}-4}(I_{n_1n_2,n_3,n_4}+\xi_2\xi_4\bar{\delta}_{n_20}\bar{\delta}_{n_40}I_{n_1,n_2-1,n_3,n_4-1}\\   \notag
&+\xi_1\xi_3\bar{\delta}_{n_10}\bar{\delta}_{n_30}I_{n_1-1,n_2,n_3-1,n_4}+\bar{\delta}_{n_10}\bar{\delta}_{n_20}\bar{\delta}_{n_30}\bar{\delta}_{n_40}\xi_1\xi_2\xi_3\xi_4I_{n_1-1,n_2-1,n_3-1,n_4-1})\\
I_{n_1,n_2,n_3,n_4}&=\sum_{k,k^\prime}\int d^2r_1d^2r_2V_C(r_1-r_2)\phi^*_{n_1k}(r_1)\phi^*_{n_2k^\prime}(r_2)\phi_{n_3k^\prime}(r_1)\phi_{n_4k}(r_2)
\end{align}
We have substituted the Landau level wavefunction from Eq.~\eqref{eq:LL wavefunction} into the above equation and defined \(\bar{\delta}_{nm} = 1 - \delta_{nm}\) for simplicity. The integral \( I_{n_1,n_2,n_3,n_4} \) can be calculated in Fourier space as follows:

\begin{align}\label{eq:fourier space}
I_{n_1,n_2,n_3,n_4}=\sum_{k,k^\prime}\int \frac{d^2q}{(2\pi)^2}\Tilde{V}_C(q)\Tilde{f}_{n_1k,n_3k^\prime}(q)\Tilde{f}_{n_2k^\prime,n_4k}(-q)
\end{align}
where \(\Tilde{f}_{nk,n^\prime k^\prime}(q)\), known as the form factor of graphene, is given by:

\begin{align}
     \Tilde{f}_{n^\prime  k^\prime,n k}(q)&   \notag =\int d^2r\phi^*_{n^\prime k^\prime}(r)\phi_{n k}(r)e^{i q\cdot r}\\   \notag
    &=\frac{1}{2^{\frac{n+n^\prime}{2}}\sqrt{n!n^\prime!\pi}}\frac{2\pi\delta(k-k^\prime+q_y)}{L_y}\\   \notag
    &\times \int dx H_{n^\prime}(x-k^\prime)H_{n}(x-k)e^{-\frac{(x-k)^2}{2}-\frac{(x-k^\prime)^2}{2}+iq_x x}\\  \notag
    &=\frac{1}{2^{\frac{n+n^\prime}{2}}\sqrt{n!n^\prime!\pi}}\frac{2\pi\delta(k-k^\prime+q_y)}{L_y} e^{-\frac{q^2}{4}+\frac{i(k+k^\prime)q_x}{2}} \int dx H_{n^\prime}(x+\frac{k-k^\prime+iq_x}{2})H_{n}(x+\frac{k^\prime-k+iq_x}{2})e^{-x^2}\\   
    &=2^{\frac{n-n^\prime}{2}}\sqrt{\frac{n^\prime!}{n!}}\frac{2\pi\delta(k-k^\prime+q_y)}{L_y}e^{-\frac{q^2}{4}+\frac{i(k+k^\prime)q_x}{2}}(\frac{q_y+iq_x}{2})^{n-n^\prime}L_{n^\prime}^{n-n^\prime}(\frac{q^2}{2}) \text{ for $n\geq n^\prime$ } \label{eq:FF of f}
\end{align}

where $L^{\alpha}_{\beta}$ is associated Laguerre polynomials.  After substituting Eq.\eqref{eq:FF of f} into Eq.\eqref{eq:fourier space}, we arrive at the following expression:

\begin{align} \label{eq:coulomb part}
&I_{n_1,n_2,n_3,n_4}=\frac{\sqrt{2}e^2}{4\pi\epsilon_0\epsilon_rl_B}J_{min(n_1,n_3),min(n_2,n_4)}^{|n_1-n_3|}\delta_{n_1+n_2,n_3+n_4}\\
&J^{a}_{n,m}=\sqrt{\frac{n!m!}{(n+a)!(m+a)!}}\int_{0}^{\infty} dx x^{2a}e^{-x^2}L_{n}^{a}(x^2)L_{m}^{a}(x^2) \\
&=\sqrt{\frac{n!m!}{(n+a)!(m+a)!}}\frac{1}{2}\frac{\Gamma(a+\frac{1}{2})\Gamma(n+\frac{1}{2})\Gamma(m+a+1)}{\Gamma(n+1)\Gamma(m+1)\Gamma(\frac{1}{2})\Gamma(a+1)} \ _3F_2(-m,a+\frac{1}{2},\frac{1}{2};-n+\frac{1}{2},a+1;1)
\end{align}
where $\ _3F_2(a_1,a_2,a_3;b_1,b_2;z)$ is generalized hypergeometric function. Therefore, the Coulomb self energy is given by:

\begin{align}\label{eq:Coulomb-fock self energy at a given landau level}
    \Sigma^{C}_{n\xi,n^\prime \xi^\prime }&=-\sum_{\substack{m, k^{\prime},k,\xi_2\\
    \xi_3}}<nk\xi ,m k^{\prime}\xi_2|V_c|m k^{\prime}\xi_3,n^\prime k \xi^\prime >\rho_{m\xi_3,m \xi_2}\\
    &=-\sum_{m,\xi_2,\xi_3}2^{\delta_{n0}+\delta_{m0}-2}\Big(I_{nm,mn}+\xi_2\xi^\prime\Bar{\delta}_{m0}\Bar{\delta}_{n0}I_{n,m-1;m,n-1}\\ \notag
    &+\xi\xi_3\Bar{\delta}_{m0}\Bar{\delta}_{n0}I_{n-1,m;m-1,n}
    +\Bar{\delta}_{m0}\Bar{\delta}_{n0}\xi\xi^\prime\xi_3\xi_2I_{n-1,m-1;m-1,n-1}\Big)\delta_{nn^\prime} \rho_{m\xi_3,m\xi_2}.
\end{align}
We have used the fact that the density matrix \(\rho_{n\xi,n^\prime\xi^\prime}\) is diagonal in the Landau level index, as a result,  the Coulomb self-energy is likewise diagonal in the Landau level index. Furthermore, the self-energy can be expressed in terms of the different components of the density matrix as follows:
\begin{align}
   \Sigma^{C}_{n\xi,n^\prime \xi^\prime } &=-2^{\delta_{n0}-1}I_{n0,0n}\rho_{0,0}-\sum_{m\geq1}2^{\delta_{n0}-2} \left( I_{nm,mn}+\Bar{\delta}_{n0}\xi\xi^\prime I_{(n-1)(m-1),(m-1)(n-1)}     \right)\cdot I \\ \notag
    &+\sum_{m\geq1}2^{\delta_{n0}-2}\Bar{\delta}_{n0}I_{n(m-1),m(n-1)}(\xi+\xi^\prime)(\rho_{m(-),m(-)}-\rho_{m(+),m(+)})\\ \notag
    & - \sum_{m\geq1}2^{\delta_{n0}-2}\left( I_{nm,mn}-\Bar{\delta}_{n0}\xi\xi^\prime I_{(n-1)(m-1),(m-1)(n-1)}  \right) (\rho_{m(+),m(-)}+\rho_{m(-),m(+)})
\end{align}
 We have used the condition \(\hat{\rho}_{n(+),n(+)} + \hat{\rho}_{n(-),n(-)} = I\) in the above expression. As observed, the large contribution from the Dirac sea causes the second term in the first line (proportional to $I$) and second lines to diverge. While these divergences do not pose a problem when calculating energy differences between competing states, determining the absolute energy requires more careful consideration. Ref.~\cite{doi:10.1142/S0217979209062104} introduces a background subtraction scheme. Let's define the density-matrix of the background \(\hat{\rho}^{bg}\) as follows:

\begin{align}\label{eq:background density}
    \hat{\rho}^{bg}_{n\xi,n^\prime\xi^\prime} =
    \begin{cases}
        \frac{1}{2}I, & \text{if } n=n^\prime= 0, \\
        \delta_{nn^\prime}I,  & \text{if } \xi=\xi^\prime=-1, \\
        0, & \text{otherwise}.
    \end{cases}
\end{align}

The divergence due to the large number of electrons in the Dirac sea can be eliminated by subtracting the background energy (arising from $\rho_{bg}$), from the total energy. For instance, using Eq.~\eqref{eq:density parametrization}, we can express the total Coulomb interaction energy in terms of the density matrix elements in the following:

\begin{align}
    E_C &= \frac{1}{2}\text{Tr}(\Sigma \rho) \\
    & = -\sum_{m,n\geq0}(I_{nm,mn}+\bar{\delta}_{n0}\bar{\delta}_{m0}I_{(n-1)(m-1),(m-1)(n-1)})\frac{(1+\sin{\theta_m}\sin{\theta_n})}{2}\\
    &-\sum_{n,m\geq1}I_{n(m-1),m(n-1)}\cos{\theta_m}\cos{\theta_n}
\end{align}
with $\theta_0 = \pi/2$. In the limit where LL mixing approach zero, $\theta_m =0$ for all $m\geq1$, this leads the total Coulomb energy reducing to the total Coulomb energy induced by the background density in Eq.~\eqref{eq:background density}.

\subsection{Self-energy of short-range interaction}

In this section, we derive the explicit formula for the self-energy, denoted as $\Sigma^{\diamond}$, resulting from short-range interactions. Considering the presence of four distinct valley-sublattice dependent terms in the anisotropic Hamiltonian, we will separate $\Sigma^{\diamond}$ into its respective components as follows:

\begin{align}\label{eq:short range self energy}
&\hat{\Sigma}^{\diamond}=\hat{\Sigma}^{zz}+\hat{\Sigma}^{\perp z}+\hat{\Sigma}^{z\perp}+\hat{\Sigma}^{\perp \perp}\\
&\hat{\Sigma}^{zz}=\hat{\Sigma}^{zz,H}-\hat{\Sigma}^{zz,F}\\
&\hat{\Sigma}^{z\perp}=\hat{\Sigma}^{z\perp,H}-\hat{\Sigma}^{z\perp,F}\\
&\hat{\Sigma}^{\perp z}=\hat{\Sigma}^{\perp z,H}-\hat{\Sigma}^{\perp z,F}\\
&\hat{\Sigma}^{\perp \perp}=\hat{\Sigma}^{\perp \perp,H}-\hat{\Sigma}^{\perp \perp,F}
\end{align}

The computation of the matrix elements is substantially simplified due to short-range delta function interactions. We will omit the detailed calculations here and directly present the results below:
\begin{align}\label{eq:g_zz sigma}
    \Sigma^{zz}_{n\xi,n^\prime \xi^\prime} &=  \sum_{m,\xi_3,\xi_2} \frac{g_{zz}}{2\pi l_B^2} 2^{\delta_{m0}+\delta_{n0}-2}\delta_{nn^\prime}\\
    &\times\Bigg( \Big(1-\bar{\delta}_{n0}\xi\xi^\prime\Big)\Big(1-\bar{\delta}_{m0}\xi_2\xi_3\Big)\text{Tr}_{sv}(\rho_{m\xi_2,m\xi_3}\tau^{z}s^0)\tau^{z}s^0-\Big(1+\xi_2\xi_3\xi\xi^\prime\bar{\delta}_{n0}\bar{\delta}_{m0} \Big)\tau^{z}s^0\rho_{m\xi_2,m\xi_3}\tau^{z}s^0 \Bigg)\\
    \Sigma^{\perp z}_{n\xi,n^\prime \xi^\prime} &=\sum_{\alpha=x,y} \sum_{m,\xi_3,\xi_2} \frac{g_{\perp z}}{2\pi l_B^2} 2^{\delta_{m0}+\delta_{n0}-2}\delta_{nn^\prime}\\ \notag
    &\times\Bigg( \Big(1-\bar{\delta}_{n0}\xi\xi^\prime\Big)\Big(1-\bar{\delta}_{m0}\xi_2\xi_3\Big)\text{Tr}_{sv}(\rho_{m\xi_2,m\xi_3}\tau^{\alpha}s^0)\tau^{\alpha}s^0-\Big(1+\xi_2\xi_3\xi\xi^\prime\bar{\delta}_{n0}\bar{\delta}_{m0} \Big)\tau^{\alpha}s^0\rho_{m\xi_2,m\xi_3}\tau^{\alpha}s^0 \Bigg)\\
    \Sigma^{z\perp}_{n\xi,n^\prime \xi^\prime} &=- \sum_{m,\xi_3,\xi_2} \frac{g_{z \perp}}{2\pi l_B^2} 2^{\delta_{m0}+\delta_{n0}-2}\delta_{nn^\prime}\Big(2\xi^\prime\xi\bar{\delta}_{n0}+2\xi_2\xi_3\bar{\delta}_{m0} \Big)\tau^{z}s^0\rho_{m\xi_2,m\xi_3}\tau^{z}s^0\\
    \Sigma^{\perp\perp}_{n\xi,n^\prime \xi^\prime} &=-\sum_{\alpha=x,y} \sum_{m,\xi_3,\xi_2} \frac{g_{\perp \perp}}{2\pi l_B^2} 2^{\delta_{m0}+\delta_{n0}-2}\delta_{nn^\prime}\Big(2\xi^\prime\xi\bar{\delta}_{n0}+2\xi_2\xi_3\bar{\delta}_{m0} \Big)\tau^{\alpha}s^0\rho_{m\xi_2,m\xi_3}\tau^{\alpha}s^0
\end{align}

There are several important features of the short-range self-energy. First, $ g_{z\perp} $ and $ g_{\perp \perp} $ contribute only to the Fock energy; the Hartree term vanishes. This is because the Landau level  wavefunctions are orthogonal between different sublattices within the same valley, and vice versa. Second, the short-range self-energy is also diagonal in the Landau level basis just like the long-range Coulomb self-energy, because the density matrix is also diagonal in the Landau level.
Another important feature is that the matrix elements of short-range self-energy  $\Sigma^{\diamond}_{n\xi,n'\xi'}$  exhibit dependence on the Landau level index solely through $\delta_{n0}$. This dependence indicates that the short-range self-energy remains constant across all Landau levels within the Dirac sea, differing with the long-range Coulomb self-energy, where matrix elements decay with increasing Landau level index, as shown in the maintex. This is a result of the delta-function approximation of the short-range interaction.


\end{widetext}

\bibliography{references}




\end{document}